\documentclass[aps,prd,twocolumn,10pt,superscriptaddress,nofootinbib,nobibnotes,longbibliography]{revtex4-1}

\usepackage{amsfonts,amsmath,amssymb,ascmac,bm,tensor}
\usepackage{fnpct} 
\usepackage{comment}
\usepackage{ifpdf}
\usepackage{slashed}
\usepackage{color}
\usepackage[mathscr]{eucal}
\usepackage[utf8]{inputenc}
\usepackage{physics}
\usepackage{cancel}
\usepackage{soul}
\usepackage{simpler-wick}
\usepackage{hyperref}
\usepackage{appendix}
\usepackage{graphicx}
\usepackage{bm}
\usepackage{tikz}
\newcommand{\D}{\mathrm{d}}

\begin{document}

\preprint{APS/123-QED}


\title{Nonperturbative stochastic inflation in perturbative dynamical background}

\author{Xiao-Quan Ye}
\email{yexiaoquan@itp.ac.cn}
\affiliation{Institute of Theoretical Physics, Chinese Academy of Sciences (CAS), Beijing 100190, China}
\affiliation{School of Physical Sciences, University of Chinese Academy of Sciences, Beijing 100049, China}

\author{Shao-Jiang Wang}
\email{schwang@itp.ac.cn (corresponding author)}
\affiliation{Institute of Theoretical Physics, Chinese Academy of Sciences (CAS), Beijing 100190, China}
\affiliation{Asia Pacific Center for Theoretical Physics (APCTP), Pohang 37673, Korea}


\begin{abstract}
Inflationary models that contain a transient ultra-slow-roll phase can exhibit strong non-perturbative dynamics, making the usual perturbative treatment of cosmological fluctuations incomplete. In such regimes, quantum diffusion and the nonlinear gravitational response of the background can both play important roles, motivating a framework that treats them systematically within quantum field theory in curved spacetime. In this work, we derive the first-order stochastic equations in quasi-de Sitter spacetime from the Schwinger-Keldysh formalism and develop a practical procedure to obtain compact stochastic equations that consistently incorporate metric perturbations via the classical Arnowitt-Deser-Misner equations. Our approach systematically captures classical non-perturbative effects while retaining the leading first-order quantum diffusion. We apply the formalism to two inflationary scenarios with an ultra-slow-roll phase, namely the Starobinsky piecewise-linear model and critical Higgs inflation. For the Starobinsky model, numerical lattice simulations validate the stochastic description and agree well with analytical results. For critical Higgs inflation, we find that the dynamics lead to a minor suppression of the power spectrum with an additional oscillation feature. Throughout, our analysis is restricted to the regime of small metric perturbations, ensuring the self-consistency of the perturbative stochastic treatment. These results establish a concrete bridge between first-principles quantum field theory in curved spacetime and the stochastic-$\delta N$ formalism for investigating non-perturbative inflationary dynamics.
\end{abstract}

\maketitle

\section{Introduction}\label{sec:introduction}

Cosmic inflation provides the most established explanation for the observed large-scale isotropy and homogeneity of our Universe~\cite{Guth:1980zm,Starobinsky:1980te,Linde:1981mu,Albrecht:1982wi}, with its predictions consistent with cosmic microwave background (CMB) data~\cite{Planck:2018jri}. The dynamics of inflation on smaller scales, however, are far less constrained by existing observational probes, including CMB observations and large-scale structure (LSS) surveys~\cite{Planck:2019kim,Chaussidon:2024qni,Chudaykin:2025vdh}. Fortunately, even large-scale observables can encode cumulative effects of small-scale dynamics through non-perturbative processes.

One such process is the formation of primordial black holes (PBHs)~\cite{Zeldovich:1967lct,Hawking:1971ei,Carr:1974nx,Carr:1975qj,Sasaki:2018dmp,Carr:2020xqk,Carr:2020gox,Carr:2021bzv,Escriva:2022duf,Carr:2023tpt} from large fluctuations generated during inflation and is thus governed by non-perturbative inflationary dynamics. Another example is the associated scalar-induced gravitational waves (SIGWs)~\cite{Baumann:2007zm, Ananda:2006af,Kohri:2018awv, Espinosa:2018eve,Domenech:2021ztg} that are largely affected by non-Gaussianity~\cite{Cai:2018dig, Unal:2018yaa, Adshead:2021hnm, Li:2023xtl, Yuan:2023ofl, Li:2025met, Perna:2024ehx,Zeng:2025cer,Caravano:2026hca}. To model these non-linear and non-perturbative effects, a common approach is to adopt the separate universe approximation (SUA)~\cite{Starobinsky:1982ee,Starobinsky:1985ibc,Sasaki:1995aw,Sasaki:1998ug,Wands:2000dp,Lyth:2003im,Rigopoulos:2003ak}, which posits that super-horizon fluctuations evolve independently on separate Hubble patches, neglecting spatial gradients. Within this framework, the stochastic-$\delta N$ formalism has been developed to capture the contributions of small-scale quantum modes once they cross the horizon and classicalize~\cite{Starobinsky:1986fx,Nambu:1987ef,Polarski:1995jg,Nambu:1988je,Kandrup:1988sc,Nambu:1989uf,Mollerach:1990zf,Salopek:1990jq,Linde:1993xx,Casini:1998wr,Finelli:2008zg,Finelli:2010sh,Clesse:2015wea,Assadullahi:2016gkk,Vennin:2016wnk,Pattison:2017mbe,Biagetti:2018pjj,Cruces:2018cvq,Ezquiaga:2019ftu,Prokopec:2019srf,Pattison:2019hef,Ballesteros:2020sre,Ando:2020fjm,Figueroa:2020jkf,Pattison:2021oen,Mishra:2023lhe,Rigopoulos:2021nhv,Tada:2021zzj,Cruces:2021iwq,Figueroa:2021zah,Cruces:2022imf,Tomberg:2023kli,Choudhury:2024jlz,Mizuguchi:2024kbl,Jackson:2024aoo,Cruces:2024pni,Sharma:2024fbr,Launay:2024qsm,Launay:2024trh,Launay:2025kef,Ahmadi:2025oon,Calderon-Figueroa:2025dto,Kawasaki:2026hnx,Murata:2026yqb}.

Over the past decades, significant effort~\cite{Hosoya:1988yz,Tsamis:2005hd,vanderMeulen:2007ah,Finelli:2008zg,Prokopec:2010be,Garbrecht:2013coa,Garbrecht:2014dca,Burgess:2014eoa,Onemli:2015pma,Burgess:2015ajz} has been devoted to establishing the theoretical background for the stochastic picture by connecting it to the underlying quantum field theory (QFT) in curved spacetime. A major part of works employs the Schwinger–Keldysh (in-in) formalism to derive the Fokker–Planck equation~\cite{PerreaultLevasseur:2013kfq,Moss:2016uix,Andersen:2021lii} and to systematically incorporate loop corrections~\cite{Kitamoto:2018dek,Gorbenko:2019rza,Kamenshchik:2021tjh,Palma:2023idj,Palma:2025oux}. Some of the works also propose the soft de Sitter effective field theory (SdEFT) to interpret the stochastic inflation as infrared EFT of QFT in curved spacetime~\cite{Cohen:2020php,Cohen:2021fzf,Cohen:2022clv,Green:2025hmo,Beneke:2026rtf}. Recently, the connection between stochastic inflation and open quantum systems has been studied extensively~\cite{Burgess:2006jn,Shandera:2017qkg,DaddiHammou:2022itk,Colas:2024lse,deKruijf:2024ufs,Salcedo:2024smn,Lopez:2025arw,Li:2025azq}. Nevertheless, most of these derivations assume an exact de Sitter background, whereas realistic inflation occurs in a quasi-de Sitter spacetime with slow-roll (SR) deviations. The first work to consider stochastic inflation in a quasi-de Sitter background is~\cite{Pinol:2020cdp}, where the authors derive stochastic equations to first order in the spatially flat gauge using a covariant Schwinger–Keldysh formalism. However, the resulting IR stochastic equations omit gradient contributions, which play an important role in inflation models with a sudden SR-USR transition~\cite{Mishra:2023lhe,Jackson:2023obv,Artigas:2024ajh,Raveendran:2025pnz}.

The alternative approach, initially developed to derive stochastic equations within the SUA framework for SR models and later extended to all orders in slow-roll parameters in leading order of gradient expansions, involves splitting the dynamical variables into ultraviolet (UV) and infrared (IR) components and performing a coarse-graining procedure~\cite{Starobinsky:1986fx,Cruces:2021iwq,Cruces:2022imf}. Recently, several works~\cite{Launay:2024qsm, Launay:2025kef, Launay:2025lnc} have provided generalized stochastic equations derived from the full set of ADM equations via a systematic linearization procedure, which incorporates metric perturbations and gradient contributions to all orders. The validity of this linearization procedure is discussed in~\cite{Launay:2024trh}.

In this paper, we extend the path-integral treatment of the QFT to derive the stochastic equations with gradient terms. Given that metric perturbations remain small in both SR and USR phases, we construct a set of stochastic equations based on the metric perturbations of the ADM equations, which is computationally cheaper than traditional numerical relativity methods. Moreover, rather than relying on perturbative methods~\cite{Briaud:2025ayt}, we employ lattice simulations to test our stochastic equations in models featuring an ultra-slow-roll (USR) phase,  which plays a crucial role in enhancing the power spectrum to generate PBHs and SIGWs.

The paper is organized as follows. After reviewing the path-integral representation of the density matrix in Sec.~\ref{sec:review}, we trace out the UV (environment) modes to derive gauge-invariant stochastic equations for first order in Sec.~\ref{sec:theory}. We then reformulate these equations in a fully non-perturbative manner. In the subsequent Sec.~\ref{sec:numerical}, we test our formalism in two representative scenarios: the Starobinsky piecewise linear model, and the critical Higgs inflation, featuring a short USR phase that enhances the small-scale power spectrum. We summarize our main results in Sec.~\ref{sec:conclusion}, and technical details are relegated to several appendices.

\section{Closed time path integral}\label{sec:review}

In this section, we review the closed time path integral formalism for density matrix~\cite{Schwinger:1960qe,Keldysh:1964ud,Calzetta:1986ey}. The density matrix evolved from the initial state $\rho_0$ at time $t_0=-\infty$ can be written as
\begin{align}
     \rho(t) = U(t,-\infty) \rho_0 U^\dagger(t,-\infty),
\end{align}
where $U(t,-\infty)$ is the unitary evolution operator from time $t_0=-\infty$ to time $t$ at which we evaluate the density matrix. 

Since the evolution operator can be expressed as the path integral,
\begin{align}
    \langle \phi_t|U(t,t_0)|\phi_0\rangle = \int_{\phi_0}^{\phi_t} D\phi e^{iS[\phi]},
\end{align}
we can project the density matrix on the field basis with boundaries 
\begin{align}
    \langle\varphi^+ |\rho(t)|\varphi^-\rangle 
    &=\int \D\phi_0\D \phi_0' \langle\varphi^+ |U(t,t_0)|\phi_0\rangle\langle\phi_0|\rho_0|\phi_0'\rangle\nonumber\\
    &\quad \times \langle\phi_0' |U^\dagger(t,t_0)| \varphi^-\rangle  \nonumber\\
    &=\int \D\phi_0\D \phi_0' \langle\phi_0|\rho_0|\phi_0'\rangle\nonumber\\
    &\quad \times\int_{\phi_0,\phi_0'}^{\phi^\pm = \varphi^\pm} D\phi^\pm  e^{i(S[\phi^+]-S[\phi^-])}\nonumber\\
    &= \mathcal{N} \int_{\Omega}^{\varphi^\pm} D\phi^\pm  e^{i(S[\phi^+]-S[\phi^-])}\nonumber\\
    &=\rho[\varphi^+,\varphi^-],
\end{align}
with $\mathcal{N}$ normalization factor of the initial state integral.

\section{The stochastic inflation from in-in formalism}\label{sec:theory}

In this section, we aim to compute the IR effective theory for long-wave modes to first order in a general single-field inflation theory. For a scalar field $\phi$ minimally coupled to gravity, the Einstein-Hilbert action is
\begin{align}
    S =& \frac{M_\mathrm{Pl}^2}{2}\int\D^4x\left(\sqrt{-g}R\right) +S_\mathrm{GHY} +\mathcal{L}_m\nonumber\\
    =&\frac{1}{2}\int\D^4 x \left(\sqrt{-g}R\right)- \int\D^3 x \sqrt{\gamma}K\nonumber\\
    &-\int\D^4x \sqrt{-g}\left(\frac{1}{2}g^{\mu\nu}\nabla_\mu\phi\nabla_\nu\phi+V(\phi)\right),
\end{align}
where we set $M_\mathrm{Pl}^2=1/(8\pi G)=1$ and introduce the Gibbons-Hawking-York boundary term to well define the variational principle~\cite{Gibbons:1977mu}. We can adopt the ADM parameterization of the metric
\begin{align}
    ds^2 = -\alpha^2 \mathrm{d}t^2 + \gamma_{ij}(\mathrm{d}x^i+\beta^i d t)(\mathrm{d}x^j+\beta^j d t),
\end{align}
where $\alpha$ and $\beta^i$ are the lapse function and shift vector respectively, $\gamma_{ij}$ is the reduced metric on constant $t$ hypersurface. In the following text, we will use the notations same as most of the ADM literature, with the covariant derivative of a four vector denoted by $;$ and the covariant derivative with respect to $\gamma_{ij}$ characterized as a vertical bar ${}_{|i}$. In ADM formalism, we can introduce the normal vector of a constant time slice $n^\mu = (\dot{\phi}/\alpha,-\beta^i/\alpha)$ to specify the direction of time. Then we can decompose the Einstein-Hilbert action into 3+1 form
\begin{align}
    S = &\frac{1}{2} \int \D t \D^3x\ \alpha \sqrt{\gamma} \left( {}^{3}R - K^2 + K_{ij}K^{ij} \right)\nonumber \\
        &+ \int \D t \D^3x\ \alpha \sqrt{\gamma} \left[ \frac{1}{2} \Pi^2 - \frac{1}{2} \partial_i \phi \partial^i \phi - V(\phi) \right],\label{eq:ADMaction}
\end{align}
with ${}^{3}R$ the Ricci scalar defined by metric $\gamma_{ij}$, and  $\Pi$ the conjugate momentum calculated as
\begin{align}
    \Pi = n^\mu\partial_\mu\phi = \frac{\dot\phi -\beta^i\partial_i\phi}{\alpha}.
\end{align}
The extrinsic curvature $K_{ij}$ is calculated explicitly as
\begin{align}
    K_{ij} = n_{i|j} = -\frac{1}{2\alpha}(\partial_t\gamma_{ij}-\beta_{i|j}-\beta_{j|i}),
\end{align}
with the trace $K\equiv \gamma^{ij}K_{ij}$.

Now we can write down the closed time integral of the density matrix for the single-field inflation system
\begin{align}
    \rho[\varphi^+,\varphi^-] = \int D\phi^\pm  Dg_{\mu\nu}^\pm  e^{i(S^+ - S^-)},
\end{align}
where 
\begin{align}
    S^\pm &= \frac{1}{2} \int \D t \D^3x \alpha^\pm \sqrt{\gamma^\pm} \left( {}^{3}R^\pm - K^{\pm2} + K^\pm_{ij}K^{\pm ij} \right)\nonumber \\
        &+ \int \D t \D^3x \alpha^\pm \sqrt{\gamma^\pm} \left[ \frac{1}{2} \Pi^{\pm2} - \frac{1}{2} \partial_i \phi^\pm \partial^{i} \phi^\pm - V(\phi^\pm) \right].
\end{align}
In principle, we can split the variables into IR and UV parts and integrate over the short-wave modes to derive the IR effective action of the system. However, the above action is highly nonlinear in the variables $\phi^+$ and $\phi^-$, which makes it challenging to solve it directly. Since the energy scale of inflation is much smaller than the Planck scale and the spacetime is quasi-de-Sitter during inflation, we can treat gravity as a classical background and adopt the perturbation expansion to simplify the density matrix of single-field inflation.

\subsection{Perturbation to first order}

After introducing the philosophy of our work, we can proceed to expand the original action in a perturbation series and derive the corresponding IR effective theory. Here, we only consider the first-order perturbations of the action, leaving the higher-order corrections to future work. 

From quantum field theory in curved spacetime, we can write down the general metric perturbations as
\begin{align}
    ds^2 =& -\alpha^2 \mathrm{d}t^2 + 2a(t)\nabla_i\beta \mathrm{d}x^i \mathrm{d}t \nonumber\\
    &+a^2(t)[(1-2\psi)\delta_{ij}+2\partial_i\partial_j E]\mathrm{d}x^i\mathrm{d}x^j,
\end{align}
where 
\begin{equation}
    \begin{aligned}
        &\alpha = A_0(1+A),\\
        &\beta = -(B_0 + B),\\
        &\psi = \psi_0 + \psi_1,\\
        &E = E_0+ E_1.\\
    \end{aligned}
\end{equation}

Due to gauge invariance, we can choose the spatially flat gauge to set $\psi =E =0$ at first order. Then we can expand the ADM action to second order
\begin{align}
    S=&S_0 + S_1 +S_2\nonumber\\
    =&\int\D t\D^3 x a^3\left[\left(-\frac{3H^2}{A_0}+\frac{\dot{\bar{\phi}}^2}{2A_0}-A_0 V(\bar{\phi})-\frac{2H\nabla^2 B_0}{aA_0}\right)\right.\nonumber\\
        &+\frac{A}{A_0}\left(3H^2-\frac{\dot{\bar{\phi}}^2}{2}-A_0^2 V(\bar{\phi})\right)\nonumber\\
        &-\frac{2H\nabla^2 B}{aA_0}
        -A_0 V'(\bar\phi)\delta\phi +
        \frac{\dot{\bar\phi}\delta\dot{\phi}}{A_0}\nonumber\\
        &-\frac{3H^2 A^2}{A_0}-\frac{2HA\nabla^2 B}{aA_0}+\frac{\delta\dot{\phi}^2-2A\dot{\bar{\phi}}\delta\dot{\phi}+A^2 \dot{\bar{\phi}}^2}{2A_0}\nonumber\\
        &- \frac{a\dot{\bar{\phi}}\partial^i B\partial_i\delta\phi}{A_0}
        -A_0\frac{\partial_i\delta\phi\partial^i\delta\phi}{2}\nonumber\\
        &\left.-A_0 A V'(\bar{\phi})\delta\phi-\frac{A_0V''(\bar{\phi})\delta\phi^2}{2}\right],
\end{align}
where we expand the scalar field as $\phi = \bar\phi +\delta\phi$, with $\partial_i\bar{\phi}=0$. Note that the second line corresponds to the zeroth-order perturbation, the third and fourth lines correspond to the first-order perturbation, and the last three lines correspond to the second-order perturbation.

If we substitute the result into the density matrix, we can derive 
\begin{align}
   \rho[\varphi^+,\varphi^-]
    =& \int D\bar{\phi}^\pm   DA_0^\pm DB_0^\pm e^{i(S_0^+ - S_0^-)}\nonumber\\
    \times& \int D\delta\phi^\pm DA^\pm D B^\pm e^{i(S_1^+-S_1^- +S_2 ^+-S_2^-)}.
\end{align}
Since the first line of the path integral is independent of the first-order variable, we can directly calculate it. For convenience, we change the variables to Schwinger-Keldysh basis, i.e., $\phi^r = (\phi^+ +\phi^-)/2$ and $\phi^a = \phi^+ -\phi^-$, and re-express the zeroth-order density matrix as
\begin{align}
    \rho[\varphi^+,\varphi^-]
    =& \int D\bar{\phi}^\pm   DA_0^\pm DB_0^\pm e^{i(S_0^+ - S_0^-)}\nonumber\\
    =&\int D\bar{\phi}^{r/a}   DA_0^{r/a} DB_0^{r/a} e^{i(S_0^+ - S_0^-)},
\end{align}
where 
\begin{align}
    S_0^+ - S_0^- 
    =& \int\D t\D^3 x a^3\left[\frac{3H^2 A_0^a}{(A_0^{r})^2}-\frac{(\dot{\bar\phi}^r)^2}{(A_0^{r})^2}\frac{A_0^a}{2}\right.\nonumber\\
    &+\frac{\dot{\bar{\phi}}^r\dot{\bar{\phi}}^a}{(A_0^{r})^2}-A_0^a V(\bar{\phi})-A_0^r  V'(\bar{\phi})\bar{\phi}^a\nonumber\\
    &+\frac{2HA_0^a\nabla^2 B_0^r}{a(A_0^{r})^2}-\frac{2HA_0^r\nabla^2 B_0^a}{a(A_0^{r})^2}\nonumber\\
    &\left.+\mathcal{O}((\bar{\phi}^a)^2,(A_0^a)^2)\right].
\end{align}
In the background order, the variables are classical, hence we retain only the leading order in $A_0^a$ and $\bar\phi^a$.
Then, we can integrate over the $A_0^a$, $B_0^a$ and $\bar\phi^a$ to derive the equation of motion at the leading order,
\begin{equation}
    \begin{aligned}
        \frac{3H^2-(\dot{\bar\phi}^r)^2/2}{(A_0^{r})^2}-V(\bar{\phi})&= 0,\\
    \frac{\D}{a^3\D t}\left(\frac{a^3\dot{\bar{\phi}}^r}{(A_0^{r})^2}\right)+V'(\bar{\phi})&=0,\\
    \nabla^2\left(\frac{2HA_0^r}{a(A_0^{r})^2}\right) &=0.
    \end{aligned}
\end{equation}
Since the physics is independent of the absolute value of the scale factor, we can freely set $A_0^r = 1$, which reduces the background equation of motion to 
\begin{equation}
    \begin{aligned}
        &\ddot{\bar{\phi}}+3H\dot{\bar{\phi}}+V'(\bar{\phi})=0,\\
        &3H^2 = \frac{\dot{\bar\phi}^2}{2}+V(\bar{\phi}).
    \end{aligned}
\end{equation}
This is exactly the Friedmann equation and the background equation of motion for $\phi$, consistent with the results from QFT in curved spacetime.

Now, the density matrix is simplified to 
\begin{align}
    \rho[\varphi^r,\varphi^a]
    =& \int D\bar{\phi}^r   DA_0^r DB_0^r \delta\left(\ddot{\bar{\phi}}+3H\dot{\bar{\phi}}+V'(\bar{\phi})\right)\nonumber\\
    &\times\delta\left(3H^2 - \frac{\dot{\bar\phi}^2}{2}-V(\bar{\phi})\right) \nonumber\\
    &\times \int D\delta\phi^{r,a} DA^{r,a} D B^{r,a} e^{i(S_1^+-S_1^- +S_2 ^+-S_2^-)}.
\end{align}
Following a similar procedure, we can consider the first-order perturbation of the action. After some integration by parts, the first-order perturbation of the action is
\begin{align}
    S_1 &=A\left(3H^2-\frac{\dot{\bar{\phi}}^2}{2}- V(\bar{\phi})\right)\nonumber\\
        &-\nabla^2\left(\frac{2H }{aA_0}\right)B
        -\left(\frac{\D}{a^3\D t}\left(a^3\dot{\bar{\phi}}\right)+V'(\bar{\phi})\right)\delta\phi,
\end{align}
which is vanishing. Consequently, the leading nontrivial perturbation of the action appears at second order. Now we move on to expand the action to the second order,
\begin{align}
S_2^+ -S_2^- 
    = &\int\D t\D^3 x a^3\left[-(3-\epsilon)H^2(2A^r A^a)\right.\nonumber\\
    &-\frac{2H}{a}\left(A^a\nabla^2 B^r+A^r\nabla^2 B^a\right)\nonumber\\
    &+\delta\dot{\phi}^r\delta\dot{\phi}^a -
    \dot{\bar{\phi}}(A^r\delta\dot{\phi}^a+A^a\delta\dot{\phi}^r)\nonumber\\
    &-a\dot{\bar{\phi}}(\partial^i B^r\partial_i\delta\phi^a+\partial^i B^a\partial_i\delta\phi^r)\nonumber\\
    &-\partial^i\delta\phi^r\partial_i\delta\phi^a-V'(\bar\phi)(A^r\delta\phi^a+A^a\delta\phi^r)\nonumber\\
    &\left.-V''(\bar\phi)\delta\phi^a\delta\phi^r\right].
\end{align}
We note that the variables $B^r$ and $B^a$ are linear in the action, hence we can integrate over it to derive the constraint equation to first order,
\begin{equation}
    \begin{aligned}
        A^r &= \frac{\dot{\bar{\phi}}\delta\phi^r}{2H},\\
        A^a &= \frac{\dot{\bar{\phi}}\delta\phi^a}{2H}.
    \end{aligned}
\end{equation}
The density matrix can be reduced to 
\begin{align}
    \rho[\varphi^r,\varphi^a]
    &= \int D\bar{\phi}^r   DA_0^r DB_0^r \delta\left(\ddot{\bar{\phi}}+3H\dot{\bar{\phi}}+V'(\bar{\phi})\right)\nonumber\\
    &\times\delta\left(3H^2 - \frac{\dot{\bar\phi}^2}{2}-V(\bar{\phi})\right) \nonumber\\
    &\times \int D\delta\phi^{r,a} DA^{r,a}\delta\left(A^{r,a} - \frac{\dot{\bar{\phi}}\delta\phi^{r,a}}{2H}\right) e^{i(S_2 ^+-S_2^-)},
\end{align}
where the exponential factor is 
\begin{align}
    S^+-S^- 
    &=\int\D t\D^3 x a^3\left[-(3-\epsilon)\epsilon H^2\delta\phi^r\delta\phi^a+\delta\dot{\phi}^r\delta\dot{\phi}^a\right.\nonumber\\
    &-\frac{\dot{\bar{\phi}^2}}{2H}\frac{\D(\delta\phi^r\delta\phi^a)}{\D t}-\partial^i\delta\phi^r\partial_i\delta\phi^a\nonumber\\
    &\left.-V''(\bar\phi)\delta\phi^a\delta\phi^r-\frac{V'(\bar\phi)\dot{\bar{\phi}}}{H}\delta\phi^r\delta\phi^a\right]\nonumber\\
    &=\int\D t\D^3 x a^3[\delta\dot{\phi}^r\delta\dot{\phi}^a -\partial^i\delta\phi^r\partial_i\delta\phi^a\nonumber\\
    &-V''(\bar\phi)\delta\phi^a\delta\phi^r
        +2H^2(3\epsilon-\epsilon^2+\epsilon\eta)\delta\phi^r\delta\phi^a].
\end{align}
We find that the substitution of constraint equations eliminates all metric perturbations, leaving the scalar-field perturbations as the dynamical variables, which is also consistent with QFT on curved spacetime.

\subsection{IR effective action to first order}

To obtain the coarse-graining effective action of the inflation system, we split the scalar perturbation into IR and UV parts by the window function,
\begin{align}
    \delta\phi_s &= \int \frac{\D^3 k}{(2\pi)^3}\left(1-W\left(\frac{k}{\sigma aH}\right)\right)\delta\phi_k(t)e^{i\vec{k}\cdot\vec{x}},\nonumber\\
    \delta\phi_e &= \int \frac{\D^3 k}{(2\pi)^3}W\left(\frac{k}{\sigma aH}\right)\delta\phi_k(t)e^{i\vec{k}\cdot\vec{x}},
\end{align}
where $W(k/\sigma aH)\equiv \theta(k-\sigma aH)$ selects modes $k$ with wavelength shorter than the coarse-graining scale and the subscripts $\delta\phi_{s/e}$ represent the system part (long wavelength) and the environment part (short wavelength), respectively.

After introducing the UV-IR split in the action, we can separate it into system, environment, and interaction parts
\begin{align}
    &S_2^+ -S_2^- 
     = S_s+S_e+S_{\mathrm{int}}\nonumber\\
        & = \int\D t\D^3 x a^3\left[\delta\dot{\phi}^r_s\delta\dot{\phi}^a_s -\partial^i\delta\phi_s^r\partial_i\delta\phi_s^a-V''(\bar\phi)\delta\phi_s^a\delta\phi_s^r\right.\nonumber\\
        & +\left. 2H^2(3\epsilon-\epsilon^2+\epsilon\eta)\delta\phi_s^r\delta\phi_s^a\right]\nonumber\\
        & + \int\D t\D^3 x a^3\left[\delta\dot{\phi}^r_e\delta\dot{\phi}^a_e -\partial^i\delta\phi_e^r\partial_i\delta\phi_e^a-V''(\bar\phi)\delta\phi_e^a\delta\phi_e^r\right.\nonumber\\
        &+ \left. 2H^2(3\epsilon-\epsilon^2+\epsilon\eta)\delta\phi_e^r\delta\phi_e^a\right]\nonumber\\
        & + 2\int\D t\D^3 x a^3\left[\delta\dot{\phi}^r_{(s}\delta\dot{\phi}^a_{e)} -\partial^i\delta\phi_{(s}^r\partial_i\delta\phi_{e)}^a\right.\nonumber\\
        &-\left.\left(V''(\bar\phi)-2H^2(3\epsilon-\epsilon^2+\epsilon\eta)\right)\delta\phi_{(s}^r\delta\phi_{e)}^a\right],
\end{align}
where the parentheses in subscripts denote $\delta\dot{\phi}^r_{(s}\delta\dot{\phi}^a_{e)} \equiv (\delta\dot{\phi}^r_{s}\delta\dot{\phi}^a_{e}+\delta\dot{\phi}^r_{e}\delta\dot{\phi}^a_{s})/2$.

To obtain the IR effective theory of the perturbation $\delta\phi$, we want to trace out the environment modes $\delta\phi_e$. We can extract the propagators from the environment part of the action,
\begin{align*}
    \begin{tikzpicture}[baseline=(current bounding box.center)]
      \draw node[above]{$r$} node[below]{$t_1$} (0,0)--(1,0) node[above]{$r$}node[below]{$t_2$};
\end{tikzpicture}
    &= \frac{1}{2}\langle 0|{\delta\phi^r_e(t_1,x_1)\delta\phi^r_e(t_2,x_2)}|0\rangle\\
     &= \frac{f_k(t_1)f_k^*(t_2)+f_k^*(t_1)f_k(t_2)}{2}\\
    &\times\theta\left(k-\sigma a_1 H_1\right)\theta\left(k-\sigma a_2 H_2\right),\\
\begin{tikzpicture}[baseline=(current bounding box.center)]
      \draw node[above]{$r$} node[below]{$t_1$} (0,0)--(1,0) node[above]{$a$}node[below]{$t_2$};
\end{tikzpicture}
    &= \theta(t_1-t_2)\langle 0|[\delta\phi^r_e(t_1,x_1),\delta\phi^a_e(t_2,x_2)]|0\rangle\\
     &= (-i)[f_k(t_1)f_k^*(t_2)-f_k^*(t_1)f_k(t_2)]\\
    &\times \theta\left(t_{1}-t_{2}\right)\theta\left(k-\sigma a_1 H_1\right)\theta\left(k-\sigma a_2 H_2\right),\\
\begin{tikzpicture}[baseline=(current bounding box.center)]
      \draw node[above]{$a$} node[below]{$t_1$} (0,0)--(1,0) node[above]{$r$}node[below]{$t_2$};
\end{tikzpicture}
    &= -\theta(t_1-t_2)\langle 0|[\delta\phi^a_e(t_1,x_1),\delta\phi^r_e(t_2,x_2)]|0\rangle\\
     &= (-i)[f_k(t_1)f_k^*(t_2)-f_k^*(t_1)f_k(t_2)]\\
    &\times \theta\left(t_{1}-t_{2}\right)\theta\left(k-\sigma a_1 H_1\right)\theta\left(k-\sigma a_2 H_2\right),
\end{align*}
and the vertices from the interaction part
\begin{align*}
    \begin{tikzpicture}[baseline=(current bounding box.center)]
    \fill(0,0) circle (2pt);
    \draw[dashed] node[below]{$t$}(0,0)--(-1,0) node[above]{$a$};
    \draw[line width=1 pt](0,0)--(1,0) node[above]{$r$};
\end{tikzpicture}\ \ &= a^3 \delta\dot{\phi }_s^a \partial_t-(ak^2 +a^3(V''(\bar\phi)\\
&-2H^2(3\epsilon-\epsilon^2+\epsilon\eta)))\delta\phi_s^a ,\\
    \begin{tikzpicture}[baseline=(current bounding box.center)]
    \fill(0,0) circle (2pt);
    \draw[dashed] node[below]{$t$}(0,0)--(-1,0) node[above]{$r$};
    \draw[line width=1 pt](0,0)--(1,0) node[above]{$a$};
\end{tikzpicture}\ \ &= a^3 \delta\dot{\phi }_s^r \partial_t-(ak^2 +a^3(V''(\bar\phi)\\
&-2H^2(3\epsilon-\epsilon^2+\epsilon\eta)))\delta\phi_s^r ,
\end{align*}
where $f_k(t)$ is the solution to Mukhanov-Sasaki equation for $\delta\phi$,
\begin{align}
    \ddot{f}_k+3H\dot{f}_k +\left(\frac{k^2}{a^2}+V''(\bar\phi)-2H^2(3\epsilon-\epsilon^2+\epsilon\eta)\right)f_k=0.
\end{align}\label{eq:MSequationfordeltaphi}
Here, $\delta\phi_e$ corresponds to solid line and $\delta\phi_s$ corresponds to dashed line.

By considering all the possible Feynman diagrams composed of the Feynman rules above, we can write down the IR effective action,
\begin{align}
    \mathcal{L}_{\mathrm{eff}} &= a^{3}\delta\dot{\phi}_{s}^r\delta\dot{\phi}_{s}^a-a\partial_{i}\delta\phi_{s}^r\partial^{i}\delta\phi_{s}^a\nonumber\\
        &-\left(V''(\bar\phi)-2H^2(3\epsilon-\epsilon^2+\epsilon\eta)\right)\delta\phi_s^a\delta\phi_s^r\nonumber\\
        &+\left(C_1\delta\dot{\phi}_{s}^a\delta\dot{\phi}_{s}^a+C_2\delta\dot{\phi}_{s}^a\delta\phi_{s}^a+C_3\delta\phi_{s}^a\delta\phi_{s}^a\nonumber\right.\\
    &+ C_4\delta\dot{\phi}_{s}^r\delta\dot{\phi}_{s}^a + C_5\delta\dot{\phi}_{s}^r\delta\phi_{s}^a\nonumber\\
    &+\left. C_6\delta\phi_{s}^r\delta\dot{\phi}_{s}^a + C_7\delta\phi_{s}^r\delta\phi_{s}^a\right),
\end{align}
where effective coefficients are calculated by the following Feynman diagrams:
\begin{equation*}
        \begin{tikzpicture}[baseline=(current bounding box.center)]
        \fill (-0.5,0)circle(2pt);
    \fill (0.5,0)circle(2pt);
        \draw[dashed] (-1.5,0) node[above]{$a$}--(-0.5,0) node[below]{$t_1$};
        \draw  (-0.5,0) node[above]{$r$}--(0.5,0) node[above]{$r$};
        \draw[dashed](1.5,0) node[above]{$a$}--(0.5,0) node[below]{$t_2$};
    \end{tikzpicture}\ + 2\times
    \begin{tikzpicture}[baseline=(current bounding box.center)]
    \fill (-0.5,0)circle(2pt);
    \fill (0.5,0)circle(2pt);
    \node[above]at(-1.5,0){$a$};
        \draw[dashed] (-1.5,0)--(-0.5,0) node[below]{$t_1$};
        \draw  (-0.5,0)node[above]{$r$}--(0.5,0) node[above]{$a$};
        \draw[dashed](1.5,0)--(0.5,0) node[below]{$t_2$};
         \node[above] at(1.5,0){$r$};
    \end{tikzpicture}.
\end{equation*}
The coefficients are
\begin{align}
    &C_1=\frac{1}{12\pi^{2}}\frac{\D\left(\sigma aH\right)^3}{\D t} |f_k(t)|^2,\nonumber\\
    &C_2 = \frac{1}{6\pi^{2}}\frac{\D\left(\sigma aH\right)^3}{\D t} \frac{\partial| f_k(t)|^2}{\partial t},\nonumber\\
    &C_3 =\frac{1}{12\pi^{2}}\frac{\D\left(\sigma aH\right)^3}{\D t} \frac{\partial^2| f_k(t)|^2}{\partial t^2},\nonumber\\
    &C_4= C_5=C_6=C_7=0.\label{eq:correlationfunctionofnoise}
\end{align}
Note that the coefficients obtained here correspond to the leading-order contribution from the environmental modes. Higher-order contributions from the potential are reserved for future work.

After tracing out the environment modes from the action, the reduced density matrix for inflation is 
\begin{align}
    \rho[\varphi^r,\varphi^a]
    &= \int D\bar{\phi}^r   DA_0^r DB_0^r \delta\left(\ddot{\bar{\phi}}+3H\dot{\bar{\phi}}+V'(\bar{\phi})\right)\nonumber\\
    &\times\delta\left(3H^2 - \frac{\dot{\bar\phi}^2}{2}-V(\bar{\phi})\right)\delta\left(A^{r,a} - \frac{\dot{\bar{\phi}}\delta\phi^{r,a}}{2H}\right) \nonumber\\
    &\times \int D\delta\phi^{r,a} DA^{r,a} e^{i S_\mathrm{eff}},
\end{align}
where the IR effective action is
\begin{align}
    S_\mathrm{eff}&=\int \D t\D^3x \left[a^{3}\delta\dot{\phi}_{s}^r\delta\dot{\phi}_{s}^a-a\partial_{i}\delta\phi_{s}^r\partial^{i}\delta\phi_{s}^a\right.\nonumber\\
    &-\left(V''(\bar\phi)-2H^2(3\epsilon-\epsilon^2+\epsilon\eta)\right)\delta\phi_s^a\delta\phi_s^r\nonumber\\
    &\left.+\left(C_1\delta\dot{\phi}_{s}^a\delta\dot{\phi}_{s}^a+C_2\delta\dot{\phi}_{s}^a\delta\phi_{s}^a+C_3\delta\phi_{s}^a\delta\phi_{s}^a\right)\right].
\end{align}

\subsection{The stochastic equations to first order}

With the reduced density matrix at hand, we intend to recover the stochastic equations of motion for the IR inflation system. After changing the time variable to $e$-folding number $N\equiv \int H\D t$, the exponential factor of the density matrix is
\begin{align}
    S_\mathrm{eff} &= \int\D N\D^3 x\frac{a^{3}}{H}\left[H^2\frac{\D\delta\phi^r}{\D N}\frac{\D\delta\phi^a}{\D N}-\partial_{i}\delta\phi^r\partial^{i}\delta\phi^a\right.\nonumber\\
    &-\left.\left(V''(\bar\phi)-2H^2(3\epsilon-\epsilon^2+\epsilon\eta)\right)\delta\phi^r\delta\phi^a\right]\nonumber\\
    &+\left(H C_1\frac{\D\delta\phi^a}{\D N}\frac{\D\delta\phi^a}{\D N}+C_2\frac{\D\delta\phi^a}{\D N}\delta\phi^a+\frac{C_3}{H}\delta\phi^a\delta\phi^a\right).
\end{align}
Note that we will leave out the subscripts $s$ in the following content for simplicity.

Since the classical effect is encoded in the r-basis, we can transform the path integral variables from $\delta\phi^{r/a}$ to $\Pi^a\equiv a^3 H\D\delta\phi^a/\D N$ and $\Phi^a\equiv a^3 H\delta\phi^a$ and interpret $\delta\phi^r/(\D\delta\phi^r/\D N)$ as the conjugate momenta. By integrating by parts, the effective action can be re-expressed as
\begin{align}
    S_{\mathrm{eff}} &= \int\D N\D^3 x\left[-\frac{\D}{\D N}\left(a^3 H\frac{\D\delta\phi^r}{\D N}\right)\delta\phi^a-\frac{a^3}{H} \partial_{i}\delta\phi^r\partial^{i}\delta\phi^a\right.\nonumber\\
    &-\frac{a^3}{H}\left(V''(\bar\phi)-2H^2(3\epsilon-\epsilon^2+\epsilon\eta)\right)\delta\phi^r\delta\phi^a\nonumber\\
    &+\left(H^2 C_1\frac{\D\delta\phi^a}{\D N}\frac{\D\delta\phi^a}{\D N}+HC_2\frac{\D\delta\phi^a}{\D N}\delta\phi^a+C_3\delta\phi^a\delta\phi^a\right)\nonumber\\
    &= -\Phi^a\left(\frac{\D}{\D N}(\frac{\D\delta\phi^r}{\D N})+(3-\epsilon)\frac{\D\delta\phi^r}{\D N}-\partial^i\partial_{i}\delta\phi^r\right.\nonumber\\
    &\left.+(V''(\bar\phi)-2H^2(3\epsilon-\epsilon^2+\epsilon\eta))\delta\phi^r\right)\nonumber\\
    &\left.+a^{-6}\left(\frac{1}{H} C_1(\Pi^a)^2+\frac{1}{H^2}C_2\Pi^a\Phi^a+\frac{1}{H^3}C_3(\Phi^a)^2\right)\right].
\end{align}
To recover the stochastic equations, we can define $\delta\pi^r \equiv \D\delta\phi^r/\D N$ and include an extra term linear in $\Pi^a$, which modifies the effective action
\begin{align}
    S_{\mathrm{eff}} &= \int\D N\D^3 x\left[\Pi^a(\frac{\D\delta\phi^r}{\D N}-\delta\pi^r)\right.\nonumber\\
    &-\Phi^a\left(\frac{\D}{\D N}(\delta\pi^r)+(3-\epsilon)\delta\pi^r-\partial^i\partial_{i}\frac{\delta\phi^r}{H^2}\right.\nonumber\\
    &+\left.\left(V''(\bar\phi)-2H^2(3\epsilon-\epsilon^2+\epsilon\eta)\right)\frac{\delta\phi^r}{H^2}\right)\nonumber\\
    &+\left. a^{-3}\left(\frac{1}{H} C_1(\Pi^a)^2+\frac{1}{H^2}C_2\Pi^a\Phi^a+\frac{1}{H^3}C_3(\Phi^a)^2\right)\right].
\end{align}
The bilinear terms of $\Pi^a$ and $\Phi^a$ can be linearized by introducing the auxiliary fields $\xi = 
\begin{pmatrix}
 \xi_\phi\\
 \xi_\pi
\end{pmatrix}$.

After using the identity  
\begin{equation*}
    e^{-i\int\D^4 x A\frac{C}{2}A^\dagger}=N\int\mathcal{D}\xi e^{i\int\D^4 x \frac{1}{2}\xi C^{-1}\xi^\dagger+ A\xi^\dagger},
\end{equation*}
where we set $A = 
\begin{pmatrix}
		\Phi^a\\
		\Pi^a
\end{pmatrix}$ and $C = 
\begin{pmatrix}
		C_1,C_2/2\\
		C_2/2,C_3
\end{pmatrix}
$, the reduced density matrix can be written as 
\begin{align}
    \rho[\varphi^a,\pi^a]
    &= \mathcal{N} \int D\bar{\phi}^r   DA_0^r DB_0^r \delta\left(\ddot{\bar{\phi}}+3H\dot{\bar{\phi}}+V'(\bar{\phi})\right)\nonumber\\
    &\times\delta\left(3H^2 - \frac{\dot{\bar\phi}^2}{2}-V(\bar{\phi})\right)\delta\left(A^{r,a} - \frac{\dot{\bar{\phi}}\delta\phi^{r,a}}{2H}\right) \nonumber\\
    &\times \int D\Pi^a D\Phi^a DA^{r,a} e^{i S'_\mathrm{eff}},
\end{align}
where 
\begin{align}
    S'_{\mathrm{eff}} &= \int\D N\D^3 x\left[\Pi^a\left(\frac{\D\delta\phi^r}{\D N}-\delta\pi^r -\xi_\phi\right)\right.\nonumber\\
    &-\Phi^a\bigg(\frac{\D \delta\pi^r}{\D N}+(3-\epsilon)\delta\pi^r-\partial^i\partial_{i}\frac{\delta\phi^r}{H^2}\nonumber\\
    &+ \left.\left.\left(V''(\bar\phi)-2H^2(3\epsilon-\epsilon^2+\epsilon\eta)\right)\frac{\delta\phi^r}{H^2}
        -\xi_\pi\right)\right]\nonumber\\
    &+ \int\D N\D^3 x \frac{1}{2}\xi C^{-1}\xi^\dagger.
\end{align}
Note that $\mathcal{N}$ is the normalization factor of the functional Gaussian integral.

After integrating over $\Phi^a$ and $\Pi^a$, the reduced density matrix is simplified to
\begin{align}
    \rho[\varphi^a,\pi^a]
    &= N \int D\bar{\phi}^r   DA_0^r DB_0^r \delta\left(\ddot{\bar{\phi}}+3H\dot{\bar{\phi}}+V'(\bar{\phi})\right)\nonumber\\
    &\times\delta\left(3H^2 - \frac{\dot{\bar\phi}^2}{2}-V(\bar{\phi})\right) \delta\left(\frac{\D\delta\phi^r}{\D N}-\delta\pi^r -\xi_\phi\right)\nonumber\\
    &\times \delta\left(\frac{\D \delta\pi^r}{\D N}+(3-\epsilon)\delta\pi^r
    -\partial^i\partial_{i}\frac{\delta\phi^r}{H^2}\right.\nonumber\\
    &+ \left.(V''(\bar\phi) -2H^2(3\epsilon-\epsilon^2+\epsilon\eta))\frac{\delta\phi^r}{H^2}-\xi_\pi\right)\nonumber\\
    &\times \int D\xi e^{i\int\D t\D^3 x \frac{1}{2}\xi C^{-1}\xi^\dagger},
\end{align}
with the two-point correlation function of $\xi$ satisfying
\begin{equation}
    \begin{aligned}
        \langle\xi_\phi(x,N)\xi_\phi(x,N')\rangle &= 2H^{-1}C_1\delta(N-N'),\\
        \langle\xi_\phi(x,N)\xi_\pi(x,N')\rangle &= H^{-2}C_2\delta(N-N'),\\
    \langle\xi_\pi(x,N)\xi_\pi(x,N')\rangle &= 2H^{-3}C_3\delta(N-N').
    \end{aligned}
\end{equation}

The stochastic equations at first order can be directly read out from the reduced density matrix
\begin{equation}
    \begin{aligned}
        \frac{\D \delta\phi}{\D N} =& \delta\pi +\xi_\phi,\\
        \frac{\D \delta\pi}{\D N} =& -(3-\epsilon)\delta\pi+\frac{\nabla^2 \delta\phi}{a^2H^2}\\
        &-\left(V''(\bar\phi)-2H^2(3\epsilon-\epsilon^2+\epsilon\eta)\right)\frac{\delta\phi}{H^2}+\xi_\pi,\\
        A = &\frac{\D \bar\phi}{\D N}\frac{\delta\phi}{2H}.
    \end{aligned}
\end{equation}
We have dropped the superscript $r$ from the stochastic equations since variables with the superscript $a$ have been integrated out. When restricted to the single-field case at first order, our results match equations (4.47) of~\cite{Pinol:2020cdp} up to an additional gradient term. Note that the above stochastic equations are gauge dependent, hence we can use the gauge invariant Mukhanov-Sasaki variable $Q\equiv \delta\phi +(\D\phi/ \D N)\psi$~\cite{Sasaki:1986hm,KodamaSasaki1984} and Bardeen potentials $\Psi_B \equiv \psi - H (aB-a^2 H (\D E/\D N))$, $\Phi_B \equiv A+H \D(aB-a^2 H \D E/\D N)/\D N$~\cite{Bardeen1980} to re-express the stochastic equations as
\begin{equation}
    \begin{aligned}
        \frac{\D Q}{\D N} &= \pi_Q +\xi_Q,\\
        \frac{\D \pi_Q}{\D N} &= -(3-\epsilon)\pi_Q+\frac{\nabla^2 Q}{a^2H^2}\\
        &-\left(V''(\bar\phi)-2H^2(3\epsilon-\epsilon^2+\epsilon\eta)\right)\frac{Q}{H^2}+\xi_{\pi_Q},\\
        \frac{\D \bar\phi}{\D N}\frac{Q}{2H} &= \Phi_B +\frac{\D\Psi_B}{\D N}+\epsilon\Psi_B,
    \end{aligned}
\end{equation}
where we set $\xi_Q = \xi_\phi$, $\xi_{\pi_Q} = \xi_\pi$. After deriving the gauge-invariant form of stochastic equations, we can expand the equations in uniform-$N$ gauge,
\begin{equation}
    \begin{aligned}
        \frac{\D Q}{\D N} &= \pi_Q +\xi_Q,\\
        \frac{\D \pi_Q}{\D N} &= -(3-\epsilon)\pi_Q+\frac{\nabla^2 Q}{a^2H^2}\\
        &-\left(V''(\bar\phi)-2H^2(3\epsilon-\epsilon^2+\epsilon\eta)\right)\frac{Q}{H^2}+\xi_{\pi_Q},\\
        \frac{\D\psi}{\D N} &= \frac{\D \bar\phi}{\D N}\frac{\delta\phi}{2H}-A.
    \end{aligned}
\end{equation}

The above uniform-$N$ gauge stochastic equations only account for leading-order perturbations and thus fail to include any higher-order contributions from the potential. In contrast, the classical ADM equations, which are computationally expensive, can incorporate nonlinear metric and potential terms. Therefore, it is natural to join the two approaches into a compact set of stochastic equations. Provided that metric perturbations are small in both SR and USR phases, we perturb the classical ADM equations with respect to $\psi$ and $E$ while keeping the potential nonlinear. This results in a closed system that is more efficient for numerical implementation,
\begin{equation}
\boxed{
\begin{aligned}
    \frac{\D\phi}{\D N} &= \pi +\xi_\phi,\\
    \frac{\D\pi}{\D N} &= -\left(3+\frac{\D \ln H_\alpha}{\D N}\right)\pi+\frac{\partial_i}{H_\alpha}\left(\frac{\partial^i\phi}{H_\alpha}\right)-\frac{V_{,\phi}}{H_\alpha^2}+\xi_{\pi},\\
    \frac{\D \ln H_\alpha}{\D N} &= -\frac{\pi^2}{2}+\frac{\nabla^2 H_\alpha^{-1}}{3a^2 H_\alpha}+\frac{2\nabla^2\psi}{3(aH_\alpha)^2}-\frac{(\nabla\phi)^2}{6(aH_\alpha)^2},\\
    \frac{\D\psi}{\D N} &= \frac{\langle\pi\rangle(\phi-\langle\phi\rangle
        )}{2}-\frac{H_\alpha-\langle H_\alpha\rangle}{\langle H_\alpha\rangle},\\
    H_\alpha^2 &= \left(3-\frac{\pi^2}{2}\right)^{-1}\left[\frac{(\nabla\phi)^2-4\nabla^2\psi}{2a^2}+V(\phi)\right],\label{eq:numerical}
\end{aligned}
}
\end{equation}
where $H_\alpha\equiv H/\alpha\approx H(1-A)$ and the stochastic correlation functions are the same as the spatially-flat gauge. The details are presented in Appendix~\ref{app:perturbation}. We stress that the noise terms are taken to be the same as the spatially flat gauge because we want to recover the gauge-invariant form of stochastic equations in uniform-$N$ gauge, which implies that we can use the correlation functions in spatially flat gauge to calculate the correlation functions of noise for the whole system.

During the slow-roll epoch, the contribution of gradient terms is negligible~\cite{Briaud:2025ayt}. We can drop the gradient terms in the stochastic equations and recover the stochastic equations in the separate universe approximation,
\begin{equation}
    \begin{aligned}
        \frac{\D\phi}{\D N} &= \pi +\xi_\phi,\\
        \frac{\D\pi}{\D N} &= -(3-\frac{\pi^2}{2})\pi-\frac{V_{,\phi}}{H_\alpha^2}+\xi_\pi,\\
        H_\alpha^2 &= \frac{V(\phi)}{3-\pi^2/2}.
    \end{aligned}
\end{equation}

\subsection{Deriving the compact stochastic equations from ADM formalism}

In our discussion above, it seems that we have guessed the compact stochastic equations from scratch. We will show that it is the first-order perturbation of the ADM equations, up to negligible higher-order metric corrections. The following derivation proceeds from the metric
\begin{align}
    ds^2 = -\alpha^2 \mathrm{d}t^2 + \gamma_{ij}(\mathrm{d}x^i-\beta^i \mathrm{d}t)(\mathrm{d}x^j-\beta^j \mathrm{d}t).
\end{align}
From the 3+1 form of Einstein-Hilbert action~\eqref{eq:ADMaction}, we can derive the following equation of motion for $\gamma_{ij}$,
\begin{align}
    & \dot{K} + \beta^i K_{,i} + \alpha^{|i}{}_{|i} - \alpha \left( {}^{3}R + K^2 \right)  - \alpha \left( \frac{1}{2} S - \frac{3}{2} \rho \right) = 0, \\
    & \dot{\tilde{K}}_{ij} + 2\alpha \tilde{K}_{il} \tilde{K}^l{}_j + \beta^k \tilde{K}_{ij|k} - 2\beta_i{}^{|k} \tilde{K}_{jk} + \alpha_{|i|j} \nonumber\\
    &-\frac{1}{3} \alpha^{|k}_{|k} \gamma_{ij} - \alpha \left( {}^{3}\tilde{R}_{ij} + \frac{1}{3} K \tilde{K}_{ij} \right) + \alpha \tilde{S}_{ij} = 0,
\end{align}
and for the field variable $\phi$,
\begin{equation}
    \frac{\dot{\Pi}+\beta^i\partial_i\Pi}{\alpha}- K\Pi -\frac{\alpha^{|i}\phi_{|i}}{\alpha}-\phi_{|i}^{|i}+\frac{\D V(\phi)}{\D\phi}=0,
\end{equation}
where the extrinsic curvature $K_{ij}$ is split into trace and traceless parts,
\begin{equation}
\begin{aligned}
    K = &\gamma^{ij}K_{ij},\\
    \tilde{K}_{ij} = &K_{ij}-\frac{1}{3}K\gamma_{ij}.
\end{aligned}
\end{equation}
From the conservation of energy and momentum tensor, we can further derive the Hamiltonian constraint and momentum constraint,
\begin{equation}
    \begin{aligned}
    {}^{3}R + \frac{2}{3} K^2 - \tilde{K}_{ij} \tilde{K}^{ij} - 2 M_\mathrm{Pl}^{-2} \rho &= 0, \\
    \tilde{K}^j{}_{i|j} - \frac{2}{3} K_{|i} - M_\mathrm{Pl}^{-2} \mathcal{P}_i &= 0,
    \end{aligned}
\end{equation}
where the components of the energy-momentum tensor are
\begin{equation}
    \begin{aligned}
        \rho&=n_{\mu}n_{\mu}T^{\mu v}=\alpha^{2}T^{00}=\frac{1}{2}\Pi^{2}+\frac{1}{2}\partial_{i}\phi\partial^{i}\phi+V(\phi),\\
        \mathcal{P}_{i}&=-n_{\mu}P_{\nu i}T^{\mu\nu}=\alpha T^{0}{}_{i}=-\Pi\partial_{i}\phi,\\
    \end{aligned}
\end{equation}
with $P_{\nu i} = g_{\nu i}+ n_\nu n_i$. The 3-dimensional stress tensor is defined as the spatial part of the energy-momentum tensor $S_{ij}\equiv T_{ij} = \nabla_i\phi\nabla_j\phi-\gamma_{ij}\mathcal{L}_m$. Similarly, we can decompose it into trace and traceless parts,
\begin{equation}
    \begin{aligned}
        S&\equiv\gamma^{ij}S_{ij}=\frac{3}{2}\Pi^2-\frac{1}{2}\partial_{i}\phi\partial^{i}\phi-3V(\phi),\\
        \tilde{S}_{ij} &= S_{ij}-\frac{1}{3}\gamma_{ij}S = \frac{1}{2}(\partial_i\phi\partial_j\phi-\frac{1}{3}\partial_k\phi\partial^k\phi\gamma_{ij}).
    \end{aligned}
\end{equation}

Since we only consider the first-order perturbations, we can expand the metric as
\begin{align}
    ds^2=&-\alpha^2\mathrm{d}t^2+2a\partial_i B \mathrm{d}x^i \mathrm{d}t\nonumber\\
    &+a^2\left[(1-2\psi)\delta_{ij}+2\partial_i\partial_jE\right]\mathrm{d}x^i \mathrm{d}x^j,
\end{align}
where $\alpha =1+A$ and $\phi = \bar\phi +\delta\phi$ are first-order perturbations, but we keep the unperturbed form for future convenience, while the residual gauge perturbations $B$,$\psi$, and $E$ are also at first order.
Then we can write down the expansion rate of $t=$constant hypersurfaces,
\begin{align}
    \theta=n^{\mu}{}_{;\mu}=\frac{3}{\alpha}\left(H-\dot{\psi}+\frac{1}{3} \nabla^{2} \sigma\right),
\end{align}
where $\sigma=\dot{E}- B$ is the shear potential. The expansion of the constant time hypersurface is the time integration of $\theta$,
\begin{align}
    \tilde{N}&=-\frac{1}{3}\int\D t\alpha K= \frac{1}{3} \int \theta\alpha \mathrm{d} t\nonumber\\
    &=\int H\D t -\psi+\frac{1}{3} \nabla^{2} \int \sigma \mathrm{d}t,
\end{align}
In uniform-$N$ gauge, we require the expansion of the constant time hypersurface to be constant, which is equivalent to
\begin{align}
    \psi=\frac{1}{3} \nabla^{2} \int \sigma \mathrm{d}t.
\end{align}
We can use the residual gauge freedom to choose $ B=0$ and further reduce the constraint to
 \begin{align}
     \psi = \frac{1}{3}\nabla^2E.
 \end{align}
 
We can now compute the geometric quantities of the spatial metric. The Christoffel connections are
\begin{align}
    \Gamma^i_{jk} = -\partial_j\psi \delta^i_k-\partial_k\psi\delta^i_j+\partial_i\psi\delta_{jk} +\partial_i\partial_j\partial_k E.
\end{align}
The extrinsic curvature is 
\begin{equation}
    \begin{aligned}
        K_{ij} &= -\frac{1}{2\alpha}\gamma_{ij,0}=-\frac{H}{\alpha}\gamma_{ij}+\frac{1}{\alpha}\left(\frac{1}{3}\delta_{ij}\nabla^2-\partial_i\partial_j\right)\dot{E},\\
        K &= \gamma^{ij}K_{ij} = -\frac{1}{2\alpha}\gamma^{ij}\gamma_{ij,t}=-\frac{\partial_t\sqrt{\gamma}}{\alpha\sqrt{\gamma}}=-\frac{3H}{\alpha },\\
        \tilde{K}_{ij} &= K_{ij}-\frac{1}{3}K\gamma_{ij} = \frac{1}{\alpha}\left(\frac{1}{3}\delta_{ij}\nabla^2-\partial_i\partial_j\right)\dot{E}, 
    \end{aligned}
\end{equation}
where $\sqrt{\gamma} = a^3(1-3\psi+\nabla^2 E) = a^3$. It is quite surprising that the extrinsic curvature in uniform-$N$ gauge can be evaluated without any assumptions of $\alpha$. The extrinsic curvature with the upper index can be written as
\begin{equation}
    \begin{aligned}
        K^i{}_j &= \frac{1}{\alpha}\left(-H\delta^i_j-(\partial^i\partial_j-\frac{1}{3}\partial^k\partial_k\delta^i_j)\dot{E}\right),\\
        \tilde{K}^i{}_j &= -\frac{1}{\alpha}\left(\partial^i\partial_j-\frac{1}{3}\partial^k\partial_k\delta^i_j\right)\dot{E}.
    \end{aligned}
\end{equation}
The Ricci curvature can be calculated as 
\begin{equation}
    \begin{aligned}
        {}^3R^i{}_j &= \partial_k\partial^k \psi\delta^i_j+ \partial^i\partial_j\psi, \\
        {}^3R &= 4\partial_i\partial^i\psi,\\
        \tilde{R}_{ij} &= \left(\partial_i\partial_j-\frac{1}{3}a^2 \partial_k\partial^k\delta_{ij}\right)\psi.\\
    \end{aligned}
\end{equation}
Now we can solve the Hamiltonian constraint and the momentum constraint,
\begin{subequations}
    \begin{align}
        2\partial_i\partial^i\psi+\frac{3H^2}{\alpha^2} =& \frac{1}{2}\left(\frac{\D\phi}{\D N}\right)^2 \frac{H^2}{\alpha^2}+ \frac{1}{2}\partial_i\phi\partial^i\phi+V(\phi),\label{eq:Hamiltoniancon}\\
        \alpha\frac{\D\phi}{\D N}\partial_i\phi -2\partial_i\alpha =& \frac{1}{3}\partial_i\alpha\partial_k\partial^k\frac{\D E}{\D N} +\frac{2\alpha}{3}\partial_i\partial_k\partial^k\frac{\D E}{\D N}\nonumber\\&-\partial_k\alpha\partial^k\partial_i\frac{\D E}{\D N}.\label{eq:momentumcon}
    \end{align}
\end{subequations}
The first equation~\eqref{eq:Hamiltoniancon} can be written in a way similar to the Friedmann equation,
\begin{align}
    \left(3-\frac{1}{2}\left(\frac{\D\phi}{\D N}\right)^2\right)\frac{H^2}{\alpha^2} = -2\partial_i\partial^i\psi+ \frac{1}{2}\partial_i\phi\partial^i\phi+V(\phi).
\end{align}
For the second equation~\eqref{eq:momentumcon}, we neglect the terms $\frac{1}{3}\partial_i\alpha\partial_k\partial^k (\D E/ \D N) $ and $-\partial_k\alpha\partial^k\partial_i(\D E/ \D N)$ since they contribute only at second order while our analysis is restricted to first-order perturbations. Hence, the momentum constraint can be simplified as
\begin{align}
    \frac{\D\phi}{\D N}\partial_i\phi-\frac{2\partial_i\alpha}{\alpha}=2\partial_i\frac{\D\psi}{\D N},
\end{align}
which is equivalent to 
\begin{align}
    \frac{\D\psi}{\D N} = \frac{\langle\pi\rangle(\phi-\langle\phi\rangle
        )}{2}-\frac{H_\alpha-\langle H_\alpha\rangle}{\langle H_\alpha\rangle}
\end{align}
at first order.

Following a similar procedure, we can expand the equation of motion for $\phi$,
\begin{align}
    \ddot{\phi}+3H\dot{\phi}-\frac{\dot{\alpha}}{\alpha}\dot{\phi}-\alpha\partial^i\alpha\partial_i\phi-\frac{\alpha^2\nabla^2\phi}{a^2}+\alpha^2 \frac{\D V}{\D \phi}=0,
\end{align}
which can be rewritten in $e$-folding numbers $N$ as
\begin{align}
    \frac{\D^2\phi}{\D N^2} &+\left(3+\frac{\D H}{H\D N}\right)\frac{\D\phi}{\D N}-\frac{\dot{\alpha}}{\alpha H}\frac{\D\phi}{\D N}\nonumber\\
    &-\alpha\partial^i\left(\frac{\alpha\partial_i\phi}{a^2 H^2}\right)+\frac{\alpha^2}{H^2}\frac{\D V}{\D \phi} = 0.
\end{align}
By extracting the characteristic structure $\frac{H}{\alpha}\frac{\D\phi}{\D N}$, we can rewrite the equation of motion as
\begin{align}
    \frac{\D}{\D N}\left(\frac{H}{\alpha}\frac{\D\phi}{\D N}\right)+\frac{3H}{\alpha}\frac{\D\phi}{\D N}-\partial_i\left(\frac{\alpha}{H}\partial^i\phi\right)+ \frac{\alpha}{H}\frac{\D V}{\D \phi}=0.
\end{align}
Finally, we can use Einstein equation for $\gamma_{ij}$ to close the system,
\begin{equation}
    \begin{aligned}
        \frac{\D}{\D N}\left(\frac{H}{\alpha}\right)-\frac{1}{3}\partial_i\partial^i\left(\frac{\alpha}{H}\right)-\frac{2\alpha}{3H}\partial_i\partial^i\psi\\
        +\frac{\alpha}{H}\left(\frac{1}{2}\left(\frac{H}{\alpha}\frac{\D\phi}{\D N}\right)^2+\frac{1}{6}\partial_i\phi\partial^i\phi\right)=0,\\
        -(\partial_i\partial_j-\frac{1}{3}\delta_{ij}\nabla^2)\ddot{E} + (\frac{\dot{\alpha}}{\alpha}-H)(\partial_i\partial_j-\frac{1}{3}\delta_{ij}\nabla^2)\dot{E}\\+\alpha(\partial_i\partial_j-\frac{1}{3}\delta_{ij}\nabla^2)\alpha
        -\alpha^2(\partial_i\partial_j-\frac{1}{3}\delta_{ij}\nabla^2)\psi\\
        +\frac{\alpha^2}{2}(\partial_i\phi\partial_j\phi-\frac{1}{3}a^2\delta_{ij}\partial_k\phi\partial^k\phi)=0,\\
        \frac{\D\psi}{\D N} = \frac{\langle\pi\rangle(\phi-\langle\phi\rangle
        )}{2}-\frac{H_\alpha-\langle H_\alpha\rangle}{\langle H_\alpha\rangle},\\
    \end{aligned}
\end{equation}
where we have applied the Hamiltonian constraint to simplify the first EOM. In summary, we have derived five equations from the ADM equations, with four unknown variables $H$, $\alpha$, $\psi$, and $\phi$. 
Hence, only the following four equations are required to develop a closed set of evolution equations for the system,
\begin{equation}
    \begin{aligned}
    \frac{\D}{\D N}\left(\frac{H}{\alpha}\frac{\D\phi}{\D N}\right)
    =&-\frac{3H}{\alpha}\frac{\D\phi}{\D N} + \partial_i\left(\frac{\alpha}{H}\partial^i\phi\right) - \frac{\alpha}{H}\frac{\D V}{\D \phi},\\
        \frac{\D}{\D N}\left(\frac{H}{\alpha}\right)=&\frac{1}{3}\partial_i\partial^i\left(\frac{\alpha}{H}\right)
        +\frac{2\alpha}{3H}\partial_i\partial^i\psi\\
        &-\frac{\alpha}{H}\left(\frac{1}{2}\left(\frac{H}{\alpha}\frac{\D\phi}{\D N}\right)^2+\frac{1}{6}\partial_i\phi\partial^i\phi\right),\\
        \frac{\D\psi}{\D N} =& \frac{\langle\pi\rangle(\phi-\langle\phi\rangle
        )}{2}-\frac{H_\alpha-\langle H_\alpha\rangle}{\langle H_\alpha\rangle},\\
        \frac{H^2}{\alpha^2} =& \frac{-2\partial_i\partial^i\psi+ \frac{1}{2}\partial_i\phi\partial^i\phi+V(\phi)}{\left(3-\frac{1}{2}\left(\D\phi/\D N\right)^2\right)}.
    \end{aligned}
\end{equation}
 The compact stochastic equations~\eqref{eq:numerical} can be recovered by adding the stochastic terms to the evolution functions of field variables $\phi$ and $\pi$.

\section{Numerical simulations}\label{sec:numerical}

In this section, we apply the stochastic equations obtained above to several cases of interest. We will first briefly introduce the simulation method for the stochastic systems and then compare it with the numerical solution of the Mukhanov-Sasaki equation. Since the Mukhanov-Sasaki equation is valid only at linear order, we can verify our stochastic equations by comparing the results with the numerical solution of the Mukhanov-Sasaki equation in the Starobinsky linear model, where non-linear effects are negligible. We expect that the stochastic equations can recover the peak structure of the piecewise-linear model.  Finally, we will apply the stochastic equations in a more realistic Higgs inflation model of critical type with a USR phase to manifest the nonperturbative effect.

\subsection{Numerical method}

Provided that the lattice simulation is only valid for at most 7 $e$-folding numbers, much smaller than the total $e$-folding number of inflation, we do not use the stochastic equations with gradient terms for the whole process of inflation. In this paper, we primarily focus on the structure of the power spectrum near the peak. Moreover, most inflationary models consistent with CMB observations do not include a long USR phase, implying that superhorizon modes will be frozen after inflation reenters the slow-roll phase. Therefore, in our numerical simulations, we evolve the modes after crossing the coarse-graining horizon until the onset of the final slow-roll phase.

The basic numerical simulation contains the following steps:
\begin{enumerate}
    \item When the modes are deep inside the horizon, we can use the Bunch-Davies (BD) vacuum as the initial condition of the inflaton field $\phi$ and set $\psi=0$ at the beginning, i.e., $$\delta\phi_i = \frac{ie^{ik\eta(N)}}{a(N)\sqrt{2k}}$$ and $$\delta\pi_i = \frac{e^{ik\eta(N)}}{a(N)\sqrt{2k}}\left(-i+\frac{k}{a(N)H(N)}\right),$$ where $\eta(N)$ is the conformal time defined as $$\eta = \int_{N}^{N_{\mathrm{end}}}\frac{\mathrm{d}N}{a(N)H(N)}.$$ Once the mode functions for $\delta\phi$ and $\delta\pi$ are obtained, we follow the procedure outlined in Ref.~\cite{Launay:2025kef} to construct $\delta\phi(x)$ in real space. This is done by computing the spectrum $|\delta\phi(k)|^2$ and multiplying it with the Fourier transformation of a Gaussian random sampling in real space. Finally, we inverse-Fourier transform the result back into real space to obtain the initial field fluctuations that obey the BD vacuum condition. 
    \item Since we have verified that the compact stochastic equations are equivalent to background Friedmann equations and Mukhanov-Sasaki equation at first two orders in Appendix~\ref{app:perturbation}, we perform direct numerical simulations of all modes starting from their initial conditions and stick to the $\delta N$-gauge throughout the simulation. The stochastic terms are precomputed using numerical or analytical solutions to the Mukhanov-Sasaki equation, without accounting for the backreaction of stochastic noise on the background dynamics. At each time step, we use the fourth-order Runge-Kutta method to integrate the field equations.
    \item We will terminate the lattice simulations when the comoving Hubble horizon reaches the same order as the lattice spacing $\Delta x$. Then, we can use the gauge-invariant quantity $\mathcal{R} = \psi-\delta\phi/\bar{\pi}$ to derive all kinds of correlation functions and extract the power spectrum.
\end{enumerate}

 In our simulations, we use the lattice of $256^3$ points to cover the range of $k/k_*\sim[1,10^2]$, which contains most of the interesting information. The full box size $L$ is set to match the comoving Hubble horizon $L = 1/(a_0 H_0)$, so that the first mode crossing the horizon corresponds to $k_{\mathrm{min}} = 2\pi/L$. Although we set all modes to satisfy the BD vacuum conditions at the onset of the simulation, modes with wavenumber smaller than $e^2/L$ are effectively near the horizon, invalidating the BD vacuum condition. Hence, we exclude long-wavelength modes that cross the horizon within less than 2 $e$-folds after the simulation begins.
 For the high-frequency modes, the cutoff is set to approximately $k_{\mathrm{max}} = N\pi/(e^2 L)$ because the simulation ends before these modes have had sufficient time to freeze out.

\subsection{The power spectrum}

After simulating the stochastic equations, we can obtain the gauge-invariant curvature perturbation at the final time slice as
\begin{equation}
    \mathcal{R} = \psi-\frac{\delta\phi}{\pi} = \psi-\frac{\phi-\bar{\phi}}{\pi}.
\end{equation}
Since the curvature perturbation $\mathcal{R}(x)$ is coarse-grained over the lattice scale, we can perform the discrete Fourier transformation introduced in Refs.~\cite{Caravano:2021pgc,Caravano:2024moy,Caravano:2025diq} to derive the curvature perturbation in Fourier space,
\begin{equation}
    \mathcal{R}_{\vec{n}} = \frac {\D x^3}{N ^3 }\sum_{ x_1, x_2 , x_{ 3 } } \mathcal{R}(\vec{x}) \ e ^ {-i\frac{2\pi}{ N } \vec{x}\cdot\vec{n} } ,\quad n_1,n_2,n_3 \in { 1,\cdots ,N },
\end{equation}
with the inverse Fourier transform written as
\begin{equation}
    \mathcal{R}(\vec{x})=\frac{1}{\D x^{3}}\sum_{n_1,n_2,n_3}\mathcal{R}_{\vec{n}}e^{i\frac{2\pi}{N}\vec{n}\cdot\vec{x}}.
\end{equation}
From the definition of the power spectrum, we can derive the following relation:
\begin{align}
    \langle\mathcal{R}_k\mathcal{R}_{k'}\rangle
        = (2\pi)^3 \delta^3(k+k') \frac{2\pi^2}{k^3}P_{\mathcal{R}}(k),
\end{align}
which can be recast in the discrete Fourier modes as
\begin{align}
    P_{\mathcal{R}}\left(k=\frac{2\pi n}{N}\right) &= \frac{k^3}{2\pi^2} \langle\mathcal{R}_k\mathcal{R}_{-k}\rangle\nonumber\\
        &= \frac{k^3}{2\pi^2} \frac{1}{\Delta\ln n}\sum_{|\ln m-\ln n|\leq\frac{\Delta\ln n}{2}} |\mathcal{R}(m)|^2,
\end{align}
where $\Delta\ln n$ is the binning parameter.

If we consider the discrete lattice structure~\cite{Caravano:2021pgc}, the effective wavenumber should be modified as
\begin{equation}
    k_{\mathrm{eff}} = {\frac {2} {\D x}}\sqrt{\sin^2 \left({\frac{\pi n_1 }{N}}\right) + \sin^2\left({\frac{\pi n_2}{N}}\right) + \sin^2\left({ \frac{\pi n_3}{N}}\right)} .
\end{equation}
Hence, the power spectrum can be expressed as
\begin{equation}
    P_{\mathcal{R}}(k_{\mathrm{eff}}) =  \frac{k_{\mathrm{eff}}^3}{2\pi^2 \Delta\ln n} \sum_{|\ln m-\ln n|\leq\frac{\Delta\ln n}{2}} |\mathcal{R}(m)|^2.
\end{equation}

\subsection{Starobinsky linear model}

In the following context, we test the stochastic equations~\eqref{eq:numerical} using both an idealistic (analytical) USR model and a more realistic (numerical) USR scenario. Again, we fix $M_\mathrm{Pl}^2=1/(8\pi G)=1$ and the coarse-graining scale $\sigma=1$ to retain gradient contributions. 

For simplicity, we will first consider the Starobinsky linear model, which has been extensively studied both numerically and analytically~\cite{Starobinsky:1992ts,Ivanov:1994pa,Leach:2001zf,Ahmadi:2022lsm,Martin:2011sn,Martin:2014kja,Pi:2022zxs,Jackson:2024aoo,Mizuguchi:2024kbl}. The model can be parameterized as
\begin{equation}
V(\phi) = 
    \begin{cases}
        V_0\left(1+\frac{A_+}{M_{\mathrm{Pl}}^2} \phi\right),\quad \phi\geq 0,\\
        V_0\left(1+\frac{A_-}{M_{\mathrm{Pl}}^2} \phi\right),\quad \phi< 0,
    \end{cases}
\end{equation}
where the potential is dominant by parameter $V_0$ with $A_{\pm} \leq M_{\mathrm{Pl}}$.  If $\Lambda\equiv A_+/A_- >1$, the evolution of the field $\phi$ first approaches the slow-roll attractor when $\phi >0$, then experiences the transient USR phase since the velocity of $\phi$ is much larger than the slow-roll attractor solution of the second stage of the potential. The power spectrum enhanced by the USR phase can support PBH production; see Refs.~\cite {Ivanov:1994pa,Leach:2001zf}.

In Sec.~\ref{sec:theory}, we have proved that the correlation functions of noise can be precomputed as~\eqref{eq:correlationfunctionofnoise} in spatially flat gauge, where $f_k(t)$ is the solution to the Mukhanov-Sasaki equation for $\delta\phi$. We can analytically solve the equations above to derive the analytic power spectra of $\delta\phi$ and $\delta\pi$ at the horizon crossing; the details are provided in Ref.~\cite{Pi:2022zxs},
\begin{widetext}
    \begin{equation}
    \begin{aligned}
    \mathcal{P}_{\phi\phi}(k=\sigma aH) = 
    \begin{cases}
        \frac{H^2}{4\pi^2}(1+\sigma^2), \phi>0\\
        \frac{H^2}{8\pi^2 \Lambda^{2}\sigma^{6}}
        \Bigg[(1+\sigma^{2})\left(9\beta^{6}\left(-1 +\Lambda\right)^{2} +18\beta^{4}\left(-1+\Lambda\right)^{2}\sigma^{2} +9\beta^{2}\left(-1+\Lambda\right)^{2}\sigma^{4} +2\Lambda^{2}\sigma^{6}\right)\\
        +3\beta\left(-1+\Lambda\right)\bigg[\bigg(-3\beta^{5}\left(-1+\Lambda\right)+3\left(-4+\beta\right)\beta^{4}\left(-1+\Lambda\right)\sigma^{2} \\
        +\beta\left(-3+4\beta\left(3-4\Lambda\right)+7\Lambda\right)\sigma^{4} 
        +\left(\beta\left(3-7\Lambda\right)+4\Lambda\right) \sigma^{6}\bigg)\cos\left[2\left(-1+\frac{1}{\beta}\right)\sigma\right]\\
        +2\sigma\big(3\left(-1 +\beta\right)\beta^{4}\left(-1 +\Lambda\right) + \beta^{2}\left(3 + 3\beta^{2}\left(-1 + \Lambda\right)- 4\Lambda\right)\sigma^{2}\\
        +(\Lambda+\beta\left(3-7\Lambda+\beta\left(-3+ 4\Lambda\right)\right))\sigma^{4}- \Lambda\sigma^{6}\big) \sin\left[2\left(-1+ \frac{1}{\beta}\sigma\right)\right]
        \bigg]\Bigg] ,\phi\leq 0,
    \end{cases}
\end{aligned}
\end{equation}
\begin{equation}
    \begin{aligned}
        \mathcal{P}_{\phi\pi}(k=\sigma aH) = 
        \begin{cases}
            -\frac{H^2\sigma^2}{4\pi^2}, \phi>0\\
            -\frac{H^2}{8\pi^{2}\Lambda^{2}\sigma^{4}}\left(9\beta^{6}\left(-1+\Lambda\right)^{2}+18\beta^{4}\left(-1+\Lambda\right)^{2}\sigma^{2}+9\beta^{2}\left(-1+\Lambda\right)^{2}\sigma^{4}+2\Lambda^{2}\sigma^{6}+3\beta\left(-1+\Lambda\right)\right)\\
            \bigg[\left(-3\beta^{5}\left(-1+\Lambda\right)-6\beta^{4}\left(-1+\Lambda\right)\sigma^{2} +\beta\left(-3+\beta\left(6-8\Lambda\right)+7\Lambda\right) \sigma^{4}+2\Lambda\sigma^{6}\right) \cos\left[2\left(-1 +\frac{1}{\beta}\right)\sigma\right] \\
            +\sigma\left(3\left(-2+\beta\right)\beta^{4} \left(-1+\Lambda\right)+2\beta^{2} \left(3-4\Lambda\right)\sigma^{2} + \left(\beta\left(3-7\Lambda\right)+ 2\Lambda\right)\sigma^{4}\right) 
            \sin\left[2\left(-1 +\frac{1}{\beta}\right)\sigma\right]\bigg] , \phi\leq0,
        \end{cases}
    \end{aligned}
\end{equation}
\begin{equation}
    \begin{aligned}
        \mathcal{P}_{\pi\pi}(k=\sigma aH) = 
        \begin{cases}
            \frac{H^2\sigma^4}{4\pi^2}, \phi>0\\
            \frac{H^2}{8\pi^2\Lambda^{2}\sigma^{2}}\bigg[9\beta^{6}(-1+\Lambda)^{2}+18\beta^{4}(-1+\Lambda)^{2}\sigma^{2}+9\beta^{2}(-1+\Lambda)^{2}\sigma^{4}+2\Lambda^{2}\sigma^{6}\\
            +3\beta(-1+\Lambda)\left(\beta(-3\beta^{4}(-1+\Lambda)+(-3+7\Lambda)\sigma^{4}\right)\cos\left[2\left(-1+\frac{1}{\beta}\right)\sigma\right]\\
            +2\sigma(-3\beta^{4}(-1+\Lambda)+\beta^{2}(3-4\Lambda)\sigma^{2}+\Lambda\sigma^{4})\sin\left[2\left(-1+\frac{1}{\beta}\right)\sigma\right]\bigg],\phi\leq0,
        \end{cases}
    \end{aligned}
\end{equation}
\end{widetext}
where we set $\beta = e^{N-N_*}$ and $V_0 = 3 M_{\mathrm{Pl}}^2H_0^2$. The parameter $N_*$ is the $e$-folding number at which the scalar field $\phi$ crosses the breaking point of the potential. The numerical parameters are listed in Table~\ref{tab:parameters}.
\begin{table}
\begin{ruledtabular}
\begin{tabular}{cccccc}
 $H/M_{\mathrm{Pl}}$ & $A_+/M_{\mathrm{Pl}}$ & $A_-/M_{\mathrm{Pl}}$ & $\bar{\phi}_{\mathrm{in}}/M_{\mathrm{Pl}}$ & $\bar{\pi}_{\mathrm{in}}/M_{\mathrm{Pl}}$ & $N_{\mathrm{end}}$ \\
\hline
$2.0 \times 10^{-6}$ & $0.01$ & $0.0004$ & $0.01$ & $-0.01$ & $5$ \\
\end{tabular}
\end{ruledtabular}
\caption{\label{tab:parameters}The input parameters of Starobinsky linear model.}
\end{table}

For numerical implementation, the noise terms $\xi_\phi$ and $\xi_\pi$ are correlated stochastic variables. We can equivalently express these two noises as linear combinations of two independent normalized Gaussian stochastic variables. We adopt the method detailed in Ref.~\cite{Jackson:2024aoo} to derive the relation between the stochastic noise and the Gaussian variable. Equivalently, if we find a new matrix $S$ satisfying $S^2 = \mathcal{P}$, where $\mathcal{P}$ is the matrix defined by the correlation functions of $\xi_\phi$ and $\xi_\pi$, 
\begin{equation}
    \begin{aligned}
    \mathcal{P}=
    \left(\begin{array}{cc}
         \mathcal{P}_{\phi\phi}& \mathcal{P}_{\pi\phi} \\
         \mathcal{P}_{\phi\pi}& \mathcal{P}_{\pi\pi}
    \end{array}\right),
    \end{aligned}
\end{equation}
we can verify that the following matrix equation is satisfied,
\begin{equation}
    \left(\begin{array}{c}
         \xi_\phi\\
         \xi_\pi
    \end{array}\right)=
    \left(\begin{array}{cc}{S_{\phi\phi}}&{S_{\pi\phi}}\\
    {S_{\pi\phi}}&{S_{\pi\pi}}
    \end{array}\right)
    \left(\begin{array}{c}
         \xi_1\\
         \xi_2
    \end{array}\right),
\end{equation}
where $\xi_1$ and $\xi_2$ are two independent Gaussian variables.
The elements of the matrix $S$ can be calculated by the Cayley-Hamilton theorem as
\begin{equation}
    S = \frac{\mathcal{P}+\sqrt{\det(\mathcal{P})}I}{\sqrt{Tr(\mathcal{P})+2\sqrt{\det(\mathcal{P})}}},
\end{equation}
where $\mathrm{Tr}(\mathcal{P})=\mathcal{P}_{\phi\phi}+\mathcal{P}_{\pi\pi}$ is the trace of $\mathcal{P}$ and $\det(\mathcal{P}) = \mathcal{P}_{\phi\phi}\mathcal{P}_{\pi\pi}-\mathcal{P}_{\phi\pi}^2$ is the determinant which is nonzero for $\sigma\neq 0$. 

\begin{figure}[htbp]
\includegraphics[width=0.9\linewidth]{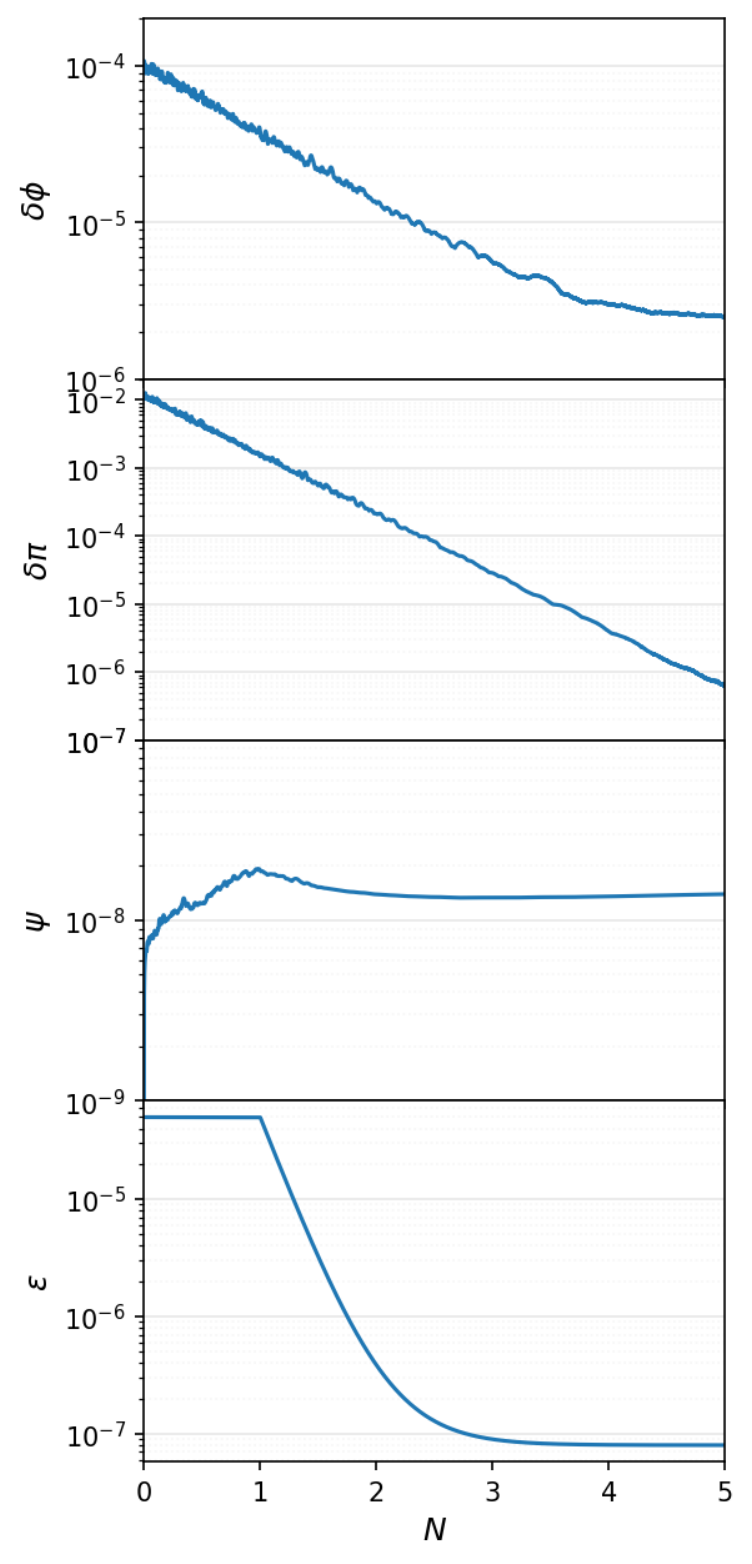}
\caption{\label{fig:starobinskysim}The maximum value of $\delta\phi$, $\delta\pi$, and $\psi$ at each simulation step extracted from the numerical simulation of the Starobinsky piecewise linear model. The last(bottom) graph shows the evolution of background value $\epsilon = \dot{H}/H$, which is calculated by the mean number of $\langle \pi^2/2\rangle$.}
\end{figure}

\begin{figure*}
\includegraphics[width=0.9\textwidth]{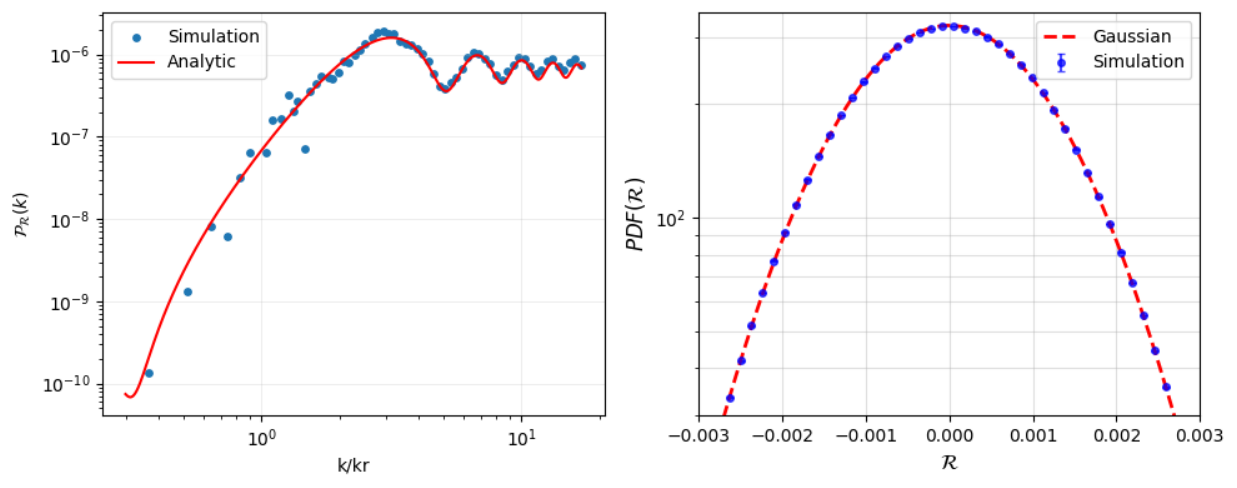}
\caption{\label{fig:Starobinskyresult}The power spectrum $\mathcal{P}_{\mathcal{R}}(k)$ (left) and probability density function(PDF) of $\mathcal{R}$ (right) of the curvature perturbation in the Starobinsky linear model. The blue dots represent the simulation result, while the red thick line represents the analytical solution to the Mukhanov–Sasaki equation in the left panel. The blue dots represent the PDF of $\mathcal{R}$ and the red dashed line represents the Gaussian fitting in the right panel.}
\end{figure*}

In Fig~\ref{fig:starobinskysim}, we show the evolution of the maximum value of the fluctuations $\delta\phi$, $\delta\pi$, and $\psi$. Although the gradient terms are included in the simulation, we find that the $\psi$ terms are strongly suppressed relative to $\delta\phi$ throughout the simulation, implying that the metric perturbations are negligible for the Starobinsky linear model. In the left panel of Fig~\ref{fig:Starobinskyresult}, we show the power spectrum computed at the end of the simulation. The analytical power spectrum of this toy model is~\cite{Pi:2022zxs}
\begin{equation}
    \begin{aligned}
    \frac{\mathcal{P}_\zeta(k)}{\mathcal{P}_{\mathrm{IR}}}
&= \frac{9(\Lambda - 1)^2}{k^6} (\sin k - k \cos k)^4\\
&+ \left[ \frac{3(\Lambda - 1)}{2k^3} \Bigl( (k^2 - 1)\sin(2k) + 2k\cos(2k) \Bigr) + \Lambda \right]^2. \\
\end{aligned}
\end{equation}
Here, we use the $P_{{\mathrm{IR}}} = (H/2\pi A_+)^2 =1.013\times10^{-9}$ and compare the numerical result with the analytical formula. Since the potential is linear, the power spectrum matches well with the analytical solution to the Mukhanov-Sasaki equation. We have shown the probability density function (PDF) of $\mathcal{R}$ at the end of the simulation in the right panel of Fig~\ref{fig:Starobinskyresult}. We have also shown the PDF of curvature perturbation with a Gaussian fitting (red dashed) for the reduced $\chi^2=1.0020$.

Though the curvature perturbation is conserved during slow-roll for each $k$ mode, the curvature perturbation simulated here is coarse-grained on the scale $k=a(N_0+N_{\mathrm{sim}})H(N_0+N_{\mathrm{sim}})$ instead of the comoving Hubble horizon scale at the end of inflation. Furthermore, the time slice is chosen as a uniform expansion slice, which differs from the uniform density slice used in curvature perturbation. Therefore, the PDF is different from the usual PDF extracted in stochastic-$\delta N$ formalism. We will further explore this problem in the next model.

\subsection{Critical Higgs inflation}

We also test the stochastic equations in a more realistic USR scenario. We use the critical Higgs inflation~\cite{Hamada:2014iga,Bezrukov:2014bra}
and modify the parameters to match the Planck result~\cite{Planck:2018jri} at approximately 55 $e$-folds. In this model, the inflaton field $\phi$ initially rolls down the potential with $\epsilon\sim\mathcal{O}(0.1)$ and then undergoes 3 $e$-folds of USR evolution before reentering the final slow-roll phase. The detailed parameters are listed in Table~\ref{tab:parametershiggs}.
\begin{table}
\begin{ruledtabular}
\begin{tabular}{ccccc}
  $\Lambda$ & $v$ & $\bar{\phi}_{\mathrm{in}}/M_{\mathrm{Pl}}$ & $\bar{\pi}_{\mathrm{in}}/M_{\mathrm{Pl}}$ & $N_{\mathrm{end}}$ \\
\hline
$9 \times 10^{-6}$ & $0.196$ & $0.455$ & $-0.217$ & $6$ \\
\end{tabular}
\end{ruledtabular}
\caption{\label{tab:parametershiggs}The input parameters of critical Higgs inflation.}
\end{table}

\begin{figure}[htbp]
\includegraphics[width=0.9\linewidth]{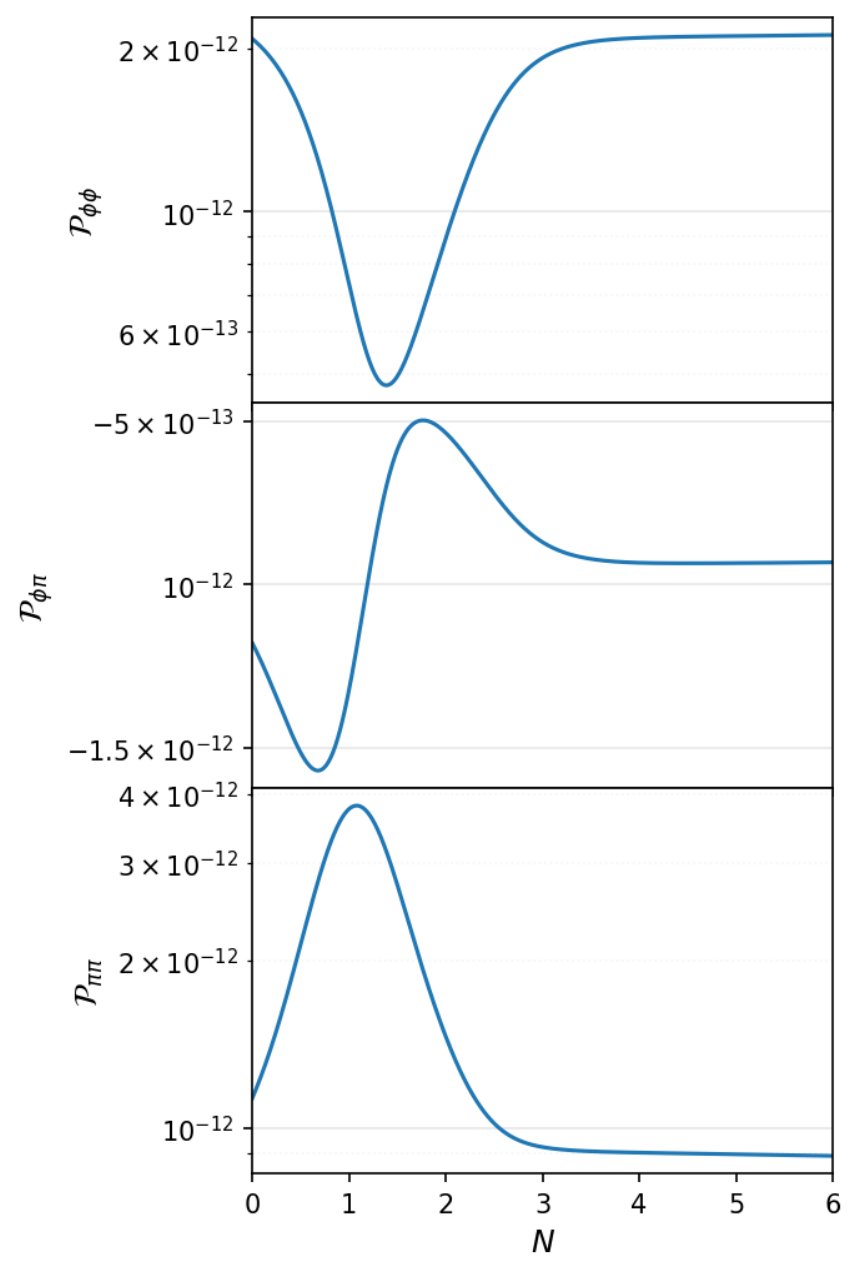}
\caption{\label{fig:higgssto}The interpolation of stochastic kicks as a function of $e$-folds for $P_{\phi\phi}$, $P_{\phi\pi}$ and $P_{\pi\pi}$ respectively.}
\end{figure}

Since the Mukhanov-Sasaki equation for critical Higgs inflation is not analytically solvable, we employ a numerical method instead. We simulate each $k$ mode from the BD vacuum state until it crosses the coarse-graining horizon. The results are then interpolated to estimate the power spectrum of an arbitrary $k$-mode that crosses the horizon during the simulation. In Fig~\ref{fig:higgssto}, we show the interpolation function of stochastic kicks.

\begin{figure}[htbp]
\includegraphics[width=0.9\linewidth]{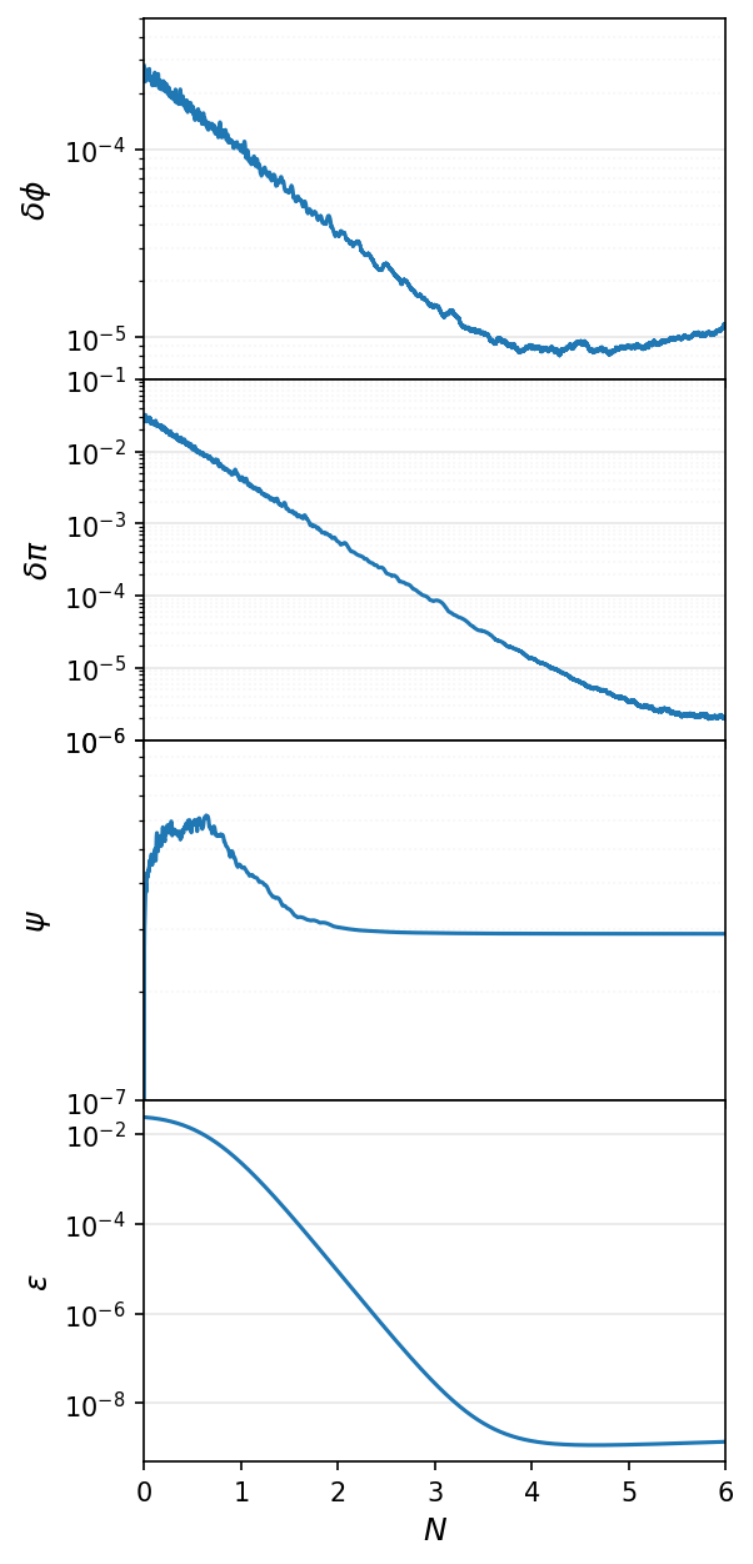}
\caption{\label{fig:higgssim}The maximum value of $\delta\phi$, $\delta\pi$, and $\psi$ at each simulation step extracted from the numerical simulation of the critical Higgs inflation. The last graph (bottom) shows the background value of $\epsilon = \dot{H}/H$, which is calculated by the mean number of $\langle \pi^2/2\rangle$.}
\end{figure}

\begin{figure*}
\includegraphics[width=0.9\textwidth]{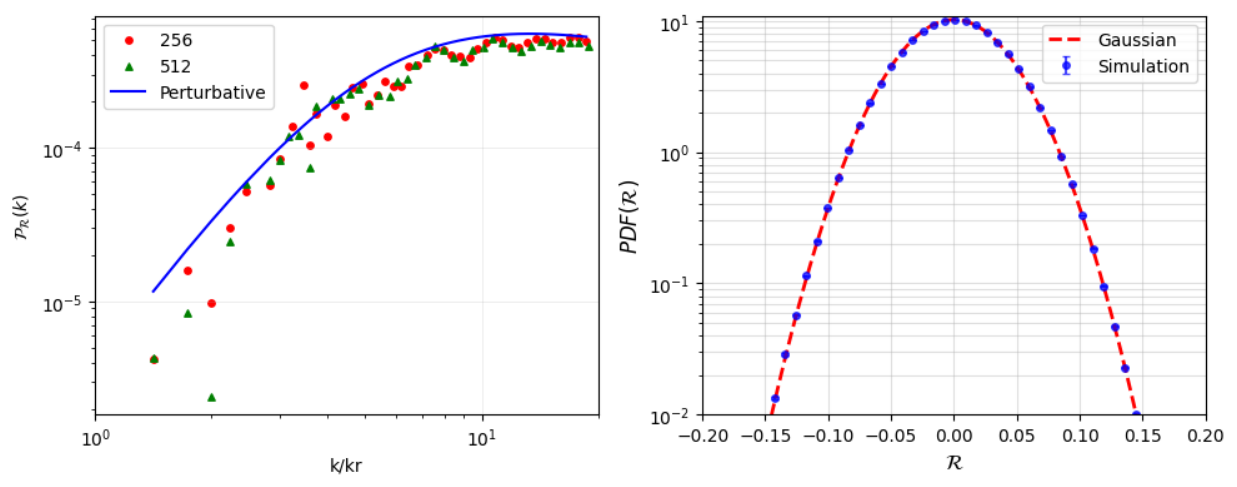}
\caption{\label{fig:higgsresult}The power spectrum $\mathcal{P}_{\mathcal{R}}(k)$ (left) and probability density function(PDF) of $\mathcal{R}$ (right) of the curvature perturbation in the critical Higgs inflation. The red dots and green triangles represent the simulation result of $256^3$ and $512^3$ grids, respectively, while the blue thick line represents the numerical solution to the Mukhanov–Sasaki equation in the left panel. The blue points with error bars represent the PDF of $\mathcal{R}$ at the end of the numerical simulation, and the red dashed line represents the Gaussian fitting in the right panel.}
\end{figure*}

In the first three panels of Fig.~\ref{fig:higgssim}, we show simulation results for the maximum value of the fluctuations $\delta\phi$, $\delta\pi$, and $\psi$. We also present the evolution of the background parameter $\epsilon$ in the bottom panel, confirming that our simulation begins at the onset of the SR-USR transition and contains the entire USR phase. We observe that the metric fluctuation $\psi$ is amplified by the initial condition $\epsilon \sim \mathcal{O}(10^{-1})$ from the slow-roll phase. Nevertheless, $\psi$ remains at least one order of magnitude smaller than $\delta\phi$, verifying our perturbative treatment. 
The left panel of Fig.~\ref{fig:higgsresult} shows the power spectrum computed at the end of the simulation. The blue thick line represents the power spectrum obtained by numerically solving the Mukhanov-Sasaki equation. The red dots and green triangles represent an average over 4 realizations for the $256^3$ and $512^3$ lattices, respectively. We observe a slightly suppressed power spectrum relative to the numerically calculated one, as reported in Ref.~\cite{Caravano:2024moy}. Furthermore, we comment on the stochastic effects on the peak of the power spectrum. Due to stochastic noises from short-wavelength modes, the peak of the power spectrum exhibits an additional oscillation structure. This behavior is attributed to non-negligible metric perturbations. According to the evolution equation for metric perturbations, the inhomogeneous field $\phi$ induces inhomogeneity in the metric perturbations. Consequently, the correlation of curvature perturbations is enhanced across different scales, giving rise to an oscillatory structure in the power spectrum. We demonstrate this effect by comparing with the same numerical simulation without metric perturbations, as shown in Fig.

Moreover, since the stochastic Eqs.~\eqref{eq:numerical} only include first-order contributions from short-wavelength noises, they fail to contain the back-reaction of these noises on the background dynamics, which may further influence the shape of the power spectrum. The oscillatory feature could be damped once such background effects are incorporated. A comprehensive treatment that accounts for significant stochastic back-reaction is left for future work.

\begin{figure*}[htbp]
\includegraphics[width=0.9\linewidth]{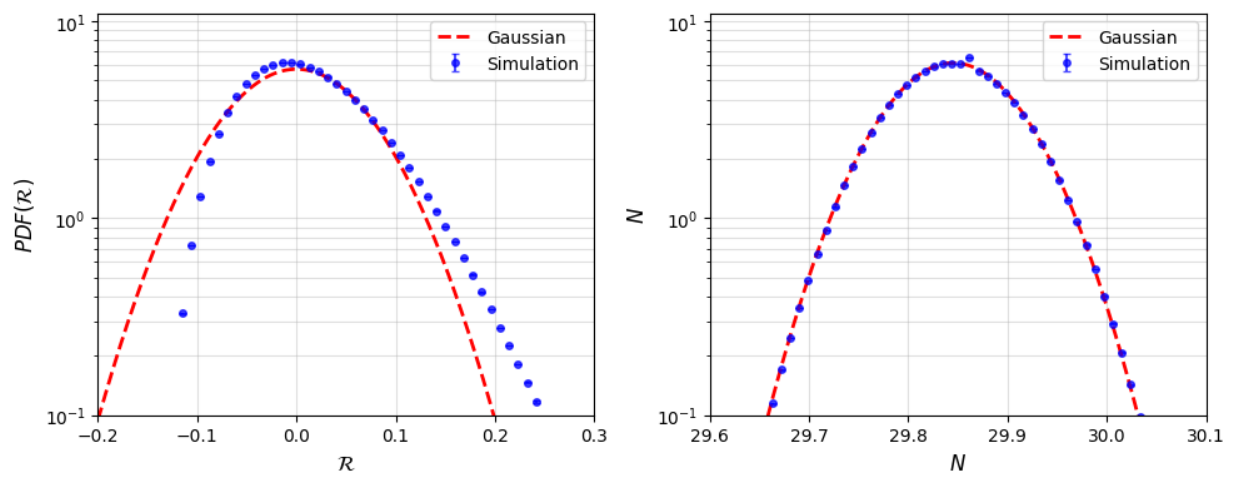}
\caption{\label{fig:higgsng}In the left panel, we show the PDF of $\mathcal{R}$ near the end of the inflation extracted on the uniform $N$ slice. The right panel shows the PDF of the total $e$-folding number $N$.}
\end{figure*}

In the right panel of Fig.~\ref{fig:higgsresult}, we present the PDF of $\mathcal{R}$ at the end of the simulation with a Gaussian fitting (red dashed) for the reduced $\chi^2=1.0020$. Provided that the transition between the USR and SR is smooth in the critical Higgs inflation, our simulation shows no pronounced non-Gaussianity by the end of the simulation, consistent with the analytical analysis in Ref.~\cite{Cai:2018dkf}.
To recover the PDF of $\mathcal{R}$, we extend the simulation time to the end of inflation. Since the gradient terms are only valid for up to 6 $e$-folds, we neglect all gradient terms in Eq.~\eqref{eq:numerical} while keeping the stochastic noises. 
The simulation continues until each lattice point reaches $\epsilon(x) = \frac{\pi^2(x)}{2}=1$. We show the results in Fig~\ref{fig:higgsng}.  We find that the shape of the PDF of curvature perturbation $\mathcal{R}$ depends on the choices of time slices. On a uniform-$N$ slice, we observe the exponential tail feature induced by large $\epsilon$ at the end of inflation. Nevertheless, the PDF of curvature perturbation on the uniform density slice remains Gaussian due to the absence of a local minimum around the inflection point of the potential.

\section{Conclusions and discussions}\label{sec:conclusion}

In this paper, we derive first-order stochastic equations in quasi-de Sitter spacetime using the Schwinger-Keldysh formalism in uniform-N gauge. By integrating out the environmental degrees of freedom and introducing auxiliary fields, we show that the noise terms in the first-order stochastic equations are described by Gaussian random noise.
We further present a method to obtain the corresponding compact stochastic equations from the classical ADM equations, thereby incorporating classical non-perturbative effects. 
These equations are then tested in two distinct USR scenarios: the well-known toy model—the Starobinsky piecewise linear model—and the more realistic critical Higgs inflation, where non-perturbative dynamics are pronounced. 
The validity of our framework is verified in the Starobinsky model by comparing numerical lattice simulations with analytical calculations. In the critical Higgs inflation, we observe a suppression of the power spectrum with an additional oscillation feature. Throughout the simulations, the metric perturbations remain small, which is consistent with our perturbative treatment of the stochastic formalism.
 
This study is limited to the first-order treatment of quantum diffusion on the small scales. In the future, a systematic incorporation of higher-order corrections will be essential for applications to models where quantum diffusion effects are strong. 
Numerically, our simulations are limited to regimes in which the stochastic noise terms remain small. An interesting future direction is to investigate scenarios with strong quantum diffusion, testing its potential impact on the shape of the power spectrum and on the full PDF of the coarse-grained curvature perturbation on the uniform-density slice at the end of inflation.
Furthermore, since our simulation framework is, in principle, capable of extracting higher-order correlation functions, such as bispectrum and trispectrum, which are not explored in detail here. We hope to apply these stochastic equations in future work to compare their predictions with those from gradient-free stochastic-$\delta N$ formalism~\cite{Vennin:2015hra}.

Finally, our present work demonstrates the feasibility of simulating the PDF of curvature perturbations via this method. Recent literature has employed advanced techniques, such as importance sampling~\cite{Jackson:2022unc,Tomberg:2022mkt,Jackson:2024aoo,Animali:2025pyf}, to efficiently capture the tail behavior of the stochastic inflation. 
Incorporating these optimization methods will improve computational efficiency, thereby enabling the accurate numerical determination of the PDF tail required for predictions of PBH abundance and the power spectrum of scalar-induced gravitational waves (SIGW).

\textbf{Note added:} 
While we are finishing writing this manuscript, a preprint on lattice simulations of inflation with metric perturbations appeared on arXiv~\cite{Saha:2026cay}. The Klein-Gordon equation for the scalar field and the Hamiltonian constraint are derived in the synchronous gauge in that paper. Here, we derive the stochastic equations~\eqref{eq:numerical} from QFT on a dynamical background while retaining the gradient contributions.

\begin{acknowledgments}
We thank Yue-Zhou Li for reading our manuscript with insightful and constructive suggestions.
This work is supported by the National Key Research and Development Program of China (Grants No. 2021YFC2203004 and No. 2021YFA0718304), and the National Natural Science Foundation of China (Grants No. 12422502, No. 12547110, No.12588101, No. 12235019, and No. 12447101). We acknowledge the use of the ITP cluster.
\end{acknowledgments}

\onecolumngrid
\appendix

\section{The calculation of coarse-graining coefficients}\label{app:coefficients}

From the Feynman rules, we can calculate the coefficients of the IR effective action
    \begin{equation}
     \begin{aligned}
        \int\D t\D^3 x_1 C_1\delta\dot{\phi}_{s}^a\delta\dot{\phi}_{s}^a
        =& \frac{1}{2}\int \mathrm{d}t_{1}\mathrm{d}t_{2}d^{3}x_{1}d^{3}x_{2}a_{1}^{3}a_{2}^{3}\delta\dot{\phi}_{s}^a(t_{1},x_{1})\delta\dot{\phi}_{s}^a(t_{2},x_{2})\\
         \times&\int\frac{d^{3}k}{(2\pi)^{3}}\sigma^{2}\frac{\D a_1H_1}{\D t_1}\frac{\D a_2H_2}{\D t_2}\delta(k-\sigma a_{1}H)\delta(k-\sigma a_{2}H)G^{rr}(t_{1},t_{2},k)e^{ik\cdot x_{12}}\\
         =&\frac{1}{2\pi^{2}}\int \mathrm{d}td^{3}x_{1}d^{3}x_{2}a^{6}\delta\dot{\phi}_{s}^a(t,x_{1})\delta\dot{\phi}_{s}^a(t,x_{2})\frac{\D(\sigma aH)^3}{6\D t} |f_k(t)|^2\frac{\sin(\sigma a H|x_{12}|)}{\sigma a H|x_{12}|};\\
         \int\D t\D^3 x_1 C_2\delta\dot{\phi}_{s}^a\delta\phi_{s}^a
         =& -\int \mathrm{d}t_1\mathrm{d}t_2d^3x_1d^3x_2a_1^3a_2^3\delta\dot{\phi}_s^a(t_1,x_1)\delta\phi_s^a(t_2,x_2)\\
         \times&\int\frac{d^3k}{(2\pi)^3}\sigma^{2}\frac{\D a_1H_1}{\D t_1}\frac{\D a_2H_2}{\D t_2}\delta(k-a_1H_1)\delta(k-a_2H_2)\partial_{t_2}G^{rr}(t_1,t_2,k)e^{ik\cdot x_{12}}\\
         =&- \frac{1}{\pi^2}\int \mathrm{d}td^{3}x_{1}d^{3}x_{2}a^{6}
         \delta\dot{\phi}_{s}^a(t,x_{1})\delta\phi_{s}^a(t,x_{2})
         \frac{\D(\sigma aH)^3}{6\D t}\frac{\partial| f_k(t)|^2}{\partial t}
         \frac{\sin(\sigma a H|x_{12}|)}{\sigma a H|x_{12}|};\\
         \int\D t\D^3 x_1 C_3\delta\phi_{s}^a\delta\phi_{s}^a
         =& \frac{1}{2}\int \mathrm{d}t_1\mathrm{d}t_2d^3x_1d^3x_2a_1^3a_2^3\delta\phi_s^a(t_1,x_1)\delta\phi_s^a(t_2,x_2)\\
         \times& \int\frac{d^3k}{(2\pi)^3}\sigma^{2}\frac{\D a_1H_1}{\D t_1}\frac{\D a_2H_2}{\D t_2}\delta(k-a_1H)\delta(k-a_2H)\partial_{t_1}\partial_{t_2}G^{rr}(t_1,t_2,k)e^{ik\cdot x_{12}}\\
         =& \frac{1}{2\pi^2} \int \mathrm{d}td^{3}x_{1}d^{3}x_{2}a^{6} \delta\phi_{s}^a(t,x_{1})\delta\phi_{s}^a(t,x_{2})\frac{\D(\sigma aH)^3}{6\D t}\frac{\partial^2| f_k(t)|^2}{\partial t^2}
         \frac{\sin(\sigma a H|x_{12}|)}{\sigma a H|x_{12}|};\\
         &C_4= C_5=C_6=C_7=0.\\
     \end{aligned}
 \end{equation}
Since we evaluate between two points that are distant by more than the
coarse-graining radius, we can take the limit $|x_1-x_2 |=0$ and the coefficients are
\begin{eqnarray}
    &&C_1=\frac{1}{12\pi^{2}}\frac{\D(\sigma aH)^3}{\D t} |f_k(t)|^2\nonumber\\
    &&C_2 = \frac{1}{6\pi^{2}}\frac{\D(\sigma aH)^3}{\D t} \frac{\partial| f_k(\eta)|^2}{\partial t}\nonumber\\
    &&C_3 =\frac{1}{12\pi^{2}}\frac{\D(\sigma aH)^3}{\D t} \frac{\partial^2| f_k(t)|^2}{\partial t^2}\nonumber\\
    &&C_4= C_5=C_6=C_7=0.
\end{eqnarray}

\section{The zeroth and first order of compact stochastic equations}\label{app:perturbation}

We can perturb the compact stochastic equations order by order to verify the relation to the background equations and to the first-order stochastic equations. Since the stochastic terms actually correspond to first-order perturbations, we drop them in the zeroth order and obtain
\begin{equation}
    \left \{
    \begin{aligned}
        \frac{\D\bar\phi}{\D N} &= \bar{\pi} ,\\
        \frac{\D\bar{\pi}}{\D N} &= -(3-\frac{\bar{\pi}^2}{2})\bar{\pi}-\frac{V_{,\phi}(\bar{\phi})}{H^2},\\
        H^2 &= \frac{V(\bar{\phi})}{3-\frac{\bar{\pi}^2}{2}},
    \end{aligned}\right.
\end{equation}
which is consistent with the background equations.

Following a similar procedure, we can derive the first-order perturbations of compact stochastic equations
\begin{subequations}
    \begin{equation}
        \frac{\D^2\delta\phi}{\D N^2}+(3-\epsilon)\frac{\D\delta\phi}{\D N} + \left(\frac{\nabla^2}{a^{2} H^2} + \frac{V_{,\phi\phi}}{H^2}\right)\delta\phi = -2\frac{V_{,\phi}}{H^2}A + \frac{\D\bar\phi}{\D N}\frac{\D A}{\D N}+\xi_\pi+\frac{\D \xi_\phi}{\D N}+(3-\epsilon)\xi_\phi,\label{eq:appBeq1}
    \end{equation}
    \begin{equation}
        3 A+\frac{\nabla^2 \psi}{a^2 H^2}=- \frac{1}{2}\left[\frac{\D\bar\phi}{\D N}\left(\frac{\D\delta\phi}{\D N}-\frac{\D\bar\phi}{\D N}A\right)+\frac{V_{,\phi}}{H^2}\delta\phi\right],\label{eq:appBeq2}
    \end{equation}
    \begin{equation}
        A+\frac{\D\psi}{\D N} = \frac{\D \bar\phi}{\D N}\frac{\delta\phi}{2H}.\label{eq:appBeq3}
    \end{equation}
\end{subequations}

Since we expect that the above equations are equivalent to stochastic equations in the Mukahanov-Sasaki variable $Q\equiv \delta\phi+\frac{\D\bar{\phi}}{\D N}\psi$, we can take the derivatives of Q's as
\begin{equation}
    \frac{\D^2Q}{\D N^2} +(3-\epsilon)\frac{\D Q}{\D N}=\frac{\D^2\delta\phi}{\D N^2}+(3-\epsilon)\frac{\D\delta\phi}{\D N}-\left(\frac{\D^2\ln H}{\D N^2}+\frac{V_{,\phi\phi}}{H^2}-2\epsilon\frac{V_{,\phi}}{H^2}\right)\psi+\left(2\frac{\D^2\bar{\phi}}{\D N^2}+(3-\epsilon)\frac{\D\bar{\phi}}{\D N}\right)\frac{\D\psi}{\D N}+\frac{\D\bar{\phi}}{\D N}\frac{\D^2 \psi}{\D N^2}.
\end{equation}
Substitute the Eq.~\eqref{eq:appBeq1} to eliminate the term $\frac{\D^2\delta\phi}{\D N^2}+(3-\epsilon)\frac{\D\delta\phi}{\D N}$
\begin{align}
    \frac{\D^2Q}{\D N^2} +(3-\epsilon)\frac{\D Q}{\D N}= &-(\frac{\nabla^2}{a^2 H^2}+\frac{V_{,\phi\phi}}{H^2})\delta\phi-2\frac{V_{,\phi}A}{H^2}+\frac{\D\bar\phi}{\D N}\frac{\D A}{\D N}+\xi_\pi+\frac{\D \xi_\phi}{\D N}+(3-\epsilon)\xi_\phi\\
    &-\left(\frac{\D^2\ln H}{\D N^2}+\frac{V_{,\phi\phi}}{H^2}-2\epsilon\frac{V_{,\phi}}{H^2}\right)\psi+\left(2\frac{\D^2\bar{\phi}}{\D N^2}+(3-\epsilon)\frac{\D\bar{\phi}}{\D N}\right)\frac{\D\psi}{\D N}+\frac{\D\bar{\phi}}{\D N}\frac{\D^2 \psi}{\D N^2}.
\end{align}
By replacing the metric perturbations $\frac{\D A}{\D N}$ with $-\frac{1}{2}(\frac{\D^2 \bar\phi}{\D N}\delta\phi+\frac{\D\bar\phi}{\D N}\frac{\D\delta\phi}{\D N})-\frac{\D^2\psi}{\D N^2}$, we find that the coefficients of $\frac{\D^2\psi}{\D N^2}$ term cancel
\begin{align}
    \frac{\D^2Q}{\D N^2} +(3-\epsilon)\frac{\D Q}{\D N}= &-\left(\frac{\nabla^2}{a^2 H^2}+\frac{V_{,\phi\phi}}{H^2}\right)\delta\phi+\frac{\D\bar{\phi}}{\D N}\left(\frac{\D^2\bar\phi}{2\D N^2} - \frac{V_{,\phi}}{H^2}\right)\delta\phi + \frac{1}{2}\left(\frac{\D\bar{\phi}}{\D N}\right)^2 \frac{\D\delta\phi}{\D N}\\
    &-\left(\frac{\D^2\ln H}{\D N^2}+\frac{V_{,\phi\phi}}{H^2}-2\epsilon\frac{V_{,\phi}}{H^2}\right)\frac{\D\bar{\phi}}{\D N}\psi-(3-\epsilon)\frac{\D\bar{\phi}}{\D N}\frac{\D\psi}{\D N}+\xi_\pi+\frac{\D \xi_\phi}{\D N}+(3-\epsilon)\xi_\phi,
\end{align}
where we have used background equation $\frac{\D^2\bar{\phi}}{\D N^2}+(3-\epsilon)\frac{\D\bar{\phi}}{\D N}+V_{,\phi}=0$ to simplify the coefficients of $\frac{\D\psi}{\D N}$.
Now we can use Eq.~\eqref{eq:appBeq2} and Eq.~\eqref{eq:appBeq3} to replace $\frac{1}{2}\left(\frac{\D\bar{\phi}}{\D N}\right)^2 \frac{\D\delta\phi}{\D N}$ with $\frac{\D\bar{\phi}}{\D N}\left[-\frac{\nabla^2}{a^2 H^2}\psi-(3-\epsilon)\left(\frac{\D\bar{\phi}}{2\D N}\delta\phi-\frac{\D \psi}{\D N}-\frac{V_{,\phi}\delta\phi}{2H^2}\right)\right]$. After some algebra, we finally derive
\begin{align}
    \frac{\D^2Q}{\D N^2} + (3-\epsilon)\frac{\D Q}{\D N}=
     \frac{\nabla^2 Q}{a^2H^2} -\left(V''(\bar\phi)-\frac{H}{a^3}\frac{\D(a^3H\bar{\pi}^2)}{\D N}\right)\frac{Q}{H^2}+\xi_\pi+\frac{\D \xi_\phi}{\D N}+(3-\epsilon)\xi_\phi.
\end{align}
Note that the first-order stochastic equations are equivalent to the following form
\begin{equation}
    \frac{\D^2Q}{\D N^2} + (3-\epsilon)\frac{\D Q}{\D N}=
     \frac{\nabla^2 Q}{a^2H^2} -\left(V''(\bar\phi)-\frac{H}{a^3}\frac{\D(a^3H\bar{\pi}^2)}{\D N}\right)\frac{Q}{H^2}+\xi_{\pi_Q}+\frac{\D \xi_Q}{\D N}+(3-\epsilon)\xi_Q.
\end{equation}
After setting $\xi_\phi = \xi_Q$ and $\xi_\pi=\xi_{\pi_Q}$, we find that the above equations are consistent at first order.

\bibliography{main}

\end{document}